\documentclass[journal]{IEEEtran}
\IEEEoverridecommandlockouts
\usepackage{cite}
\usepackage{graphicx}
\usepackage{amsmath,amssymb,amsfonts}
\usepackage{algorithmic}
\usepackage{graphicx}
\usepackage{textcomp}
\usepackage{xcolor}
\usepackage{float,color,comment}
\usepackage{amsmath}
\usepackage{caption}
\usepackage{subfigure}
\usepackage{graphics} % for pdf, bitmapped graphics files
\usepackage{epsfig} 
\usepackage{multirow}
\graphicspath{{figure/}{figure/lstm/}}
\def\BibTeX{{\rm B\kern-.05em{\sc i\kern-.025em b}\kern-.08em
    T\kern-.1667em\lower.7ex\hbox{E}\kern-.125emX}}

\usepackage{array}
\usepackage{import}
\usepackage{xifthen}
\usepackage{pdfpages}
\usepackage{transparent}

\begin{document}

\title{A Markovian Model-Driven Deep Learning Framework for Massive MIMO CSI Feedback}

\author{\IEEEauthorblockN{Zhenyu Liu, Mason del Rosario, and Zhi Ding} }
% Zhenyu should serve as author for correspondence. Mason should
% pen this paper.

\maketitle

\begin{abstract}
Forward channel state information (CSI) often plays 
a vital role in scheduling and capacity-approaching
transmission optimization for massive multiple-input multiple-output (MIMO) communication
systems. In frequency division duplex (FDD) massive MIMO systems, forwardlink CSI 
reconstruction at the transmitter
relies critically on CSI feedback from receiving nodes and must carefully weigh the tradeoff between reconstruction 
accuracy and feedback bandwidth.
Recent studies on the use of 
recurrent neural networks (RNNs) have demonstrated strong promises, though
the cost of computation and memory remains high,
for massive MIMO deployment. In this work, we exploit 
channel coherence in time to substantially improve the
feedback efficiency. Using a Markovian model,
we develop a deep convolutional neural network (CNN)-based framework MarkovNet to differentially
encode forward CSI in time to effectively improve reconstruction
accuracy. Furthermore, we explore important physical insights, including spherical normalization of input data and convolutional layers for feedback compression.
%to improve the CNN performance. 
We demonstrate substantial performance improvement
and complexity reduction over the RNN-based work by our proposed MarkovNet to recover forward CSI estimates accurately. We explore additional practical consideration in feedback quantization, and show that MarkovNet outperforms RNN-based CSI estimation
networks at a fraction of the computational cost.

\end{abstract}


\section{Introduction}

Massive MIMO wireless
interface has been identified as a critical radio technology at the physical layer 
capable of
substantially improving the bandwidth efficiency and 
delivering Gigabits/s services 
to many heterogeneous subscribers simultaneously. 
The efficacy of such massive MIMO downlink depends on the availability of accurate 
forward (downlink)
CSI estimates at the base station (BS) for 
transmission precoding. 
Given the large number of antennas in massive MIMO
and potentially broad bandwidth, such downlink CSI estimation and acquisition require
substantial amount of feedback from each subscriber user equipment (UE).
To support high mobility UEs in modern mobile wireless, 
timely feedback for time varying (i.e., fading)
CSI estimation \cite{ref:simon2011joint,ref:song2006channel}
pose critical challenges. Frequent reporting of CSI for massive MIMO
coverage would consume too much network bandwidth and
UE power.  The need for efficient CSI feedback in massive MIMO
networks strongly motivates many
research efforts aimed at downlink CSI compression, feedback, and reconstruction.

% \begin{table}[]
% \renewcommand{\arraystretch}{1.5}
% \centering
% \caption{CsiNet-LSTM parameter count and computational complexity.}
% \label{tab:num-param}
% \begin{tabular}{|c|c|c|c|}
% \hline
%                             & \# Parameters & FLOPs  \\ \hline
% \textbf{CR=$\frac {1}{4}$}  & 132.7 M         & 412.9 M \\ \hline
% \textbf{CR=$\frac {1}{8}$}  & 123.2 M         & 410.8 M \\ \hline
% \textbf{CR=$\frac {1}{16}$} & 118.5 M         & 409.8 M \\ \hline
% \textbf{CR=$\frac {1}{32}$} & 116.1 M         & 409.2 M \\ \hline
% \textbf{CR=$\frac {1}{64}$} & 115.0 M         & 409.0 M \\ \hline
% \end{tabular}
% \end{table}

The problem of
CSI feedback and reconstruction in massive MIMO has been an active research area in recent years. Traditional vector quantization and codebook-based methods reduce feedback overhead by quantizing the CSI at the UE side \cite{ref:quan1,ref:quan4,ref:quan2,ref:quan3}. However, the feedback overhead grows with the number of antennas, often requiring large amount
of uplink bandwidth or low accuracy for practical massive MIMO wireless transmission. 
Compressive sensing (CS)-based approaches exploit the sparsity channel property in some domain to lower the CSI feedback overhead \cite{ref:cs1,ref:cs2,ref:cs3}. However,  CS-based approaches often hinge on strong channel sparsity conditions not strictly 
satisfied in some domains. Moreover, iterative CS reconstruction methods 
may need a large amount of computation time to accurately recover downlink CSI
estimates.

There has been a surging wave of interest
 in applying artificial neural
networks for forward CSI estimation \cite{Chen2006,Tacspinar2010,Cheng2013,Hiray2016}. The popularity and versatility of deep learning (DL) have
motivated a number recent works that explored
deep neural networks (DNN) for downlink channel compression and recovery, particularly
for massive MIMO wireless interface. Typically, these
DNNs have utilized two prevailing architectures that are successful in
other applications. 
First, Convolutional Neural Networks (CNNs), which have demonstrated state-of-the-art performance in image processing tasks \cite{ref:He2016}, have been 
integrated in an autoencoder for CSI compression and recovery of
a single snapshot \cite{ref:csinet}. Second, Recurrent Neural Networks (RNNs) have been 
further investigated to exploit temporal CSI coherent for feedback compression
in massive MIMO systems \cite{ref:csinet-lstm,ref:Lu2019,ref:Liao2019,ref:Li2020,ref:Jang2019}. RNNs can leverage hidden states through architectures such as long short-term memory 
(LSTM) cells to exploit the effect of past inputs. 

Existing works have demonstrated that DNNs can provide
efficient CSI feedback and reconstruction 
for time-varying MIMO 
channels \cite{ref:csinet-lstm,ref:Lu2019,ref:Liao2019,ref:Li2020,ref:Jang2019}. 
However,  important issues remain unresolved in at least the following two
aspects:

\begin{enumerate}
\item \textbf{Complexity and Storage}: The number of parameters in the RNN layers
for CSI compression and reconstruction of massive MIMO systems can be
staggeringly large. For example, the RNN module can add $10^8$ additional parameters \cite{ref:csinet-lstm}, which raises storage and computation concerns. A fully connected layer-based autoencoder has been proposed for the CSI feedback in time varying channel to 
reduce the computational 
complexity and required memory \cite{ref:Jang2019} . 
However, the accuracy is less favorable in comparison to
\cite{ref:csinet-lstm}. 
While other works have investigated RNNs of
reduced size \cite{ref:Jang2019,ref:Li2020},
the least computationally expensive of these models still requires $10^7$
parameters per snapshot. Also, the networks in \cite{ref:Jang2019, ref:Li2020} suffer from the significant feedback performance drop when the compression ratio is small, since the networks have to use the same compression ratio in successive time slots and can not get the accurate prior information in the initial time slot.
Despite the reported success of ``stacked'' LSTMs, the minimum necessary depth of recurrent layers
for CSI recovery accuracy has not been
adequately evaluated. Considering the large RNN parameter count,
performance improvement should be substantial to justify the memory overhead.

% ref:Lu2019
 
% Different avenues have been investigated to alleviate the memory burden of RNNs. 
% A deep autoencoder based CSI feedback-reconstruction
% mechanism \cite{ref:Jang2019} has used the fully connected layers to 
% reduce the computational 
% complexity and required memory. 
% However, the accuracy is less favorable in comparison to
% \cite{ref:csinet-lstm}.  RecCsiNet has been proposed in \cite{ref:Lu2019} to 
% reduce the model size of LSTM-based CSI feedback-reconstruction
% while maintaining similar performance gains.  Although the model size is reduced
% to about $\frac{1}{5}$ of the model size in \cite{ref:csinet-lstm}, 
% more than $10^7$ parameters are still required per snapshot.

% \item \textbf{Feedback delay}: 
% For high feedback efficiency and recovery accuracy,  ConvlstmCsiNet  \cite{ref:Li2020}
% has been proposed 
% to exploit the underlying spatial-temporal features for the FDD MIMO system. 
% However, this improved efficiency is at the cost of feedback 
% delay since the UEs implement CSI encoding and compression for
% $T$ time slots jointly. 

\item \textbf{Physical Insight}: The
success of RNNs in areas such as video processing \cite{ref:Milan2017} and natural language processing (NLP) \cite{ref:Sutskever2014,ref:Irsoy2014} has stimulated
their applications in forward CSI feedback and reconstruction. However, 
despite the apparent similarities in terms of a time series prediction, 
the physical nature of underlying CSI in massive MIMO
is considerably different from those in video contents and image contents.
Leveraging domain knowledge and physical characteristics 
on mobile wireless channels can be beneficial.
For example, LTE frames (subframes) occupy 10ms (1ms) of airtime 
and permit CSI feedback intervals that are integer multiples of either. 
DNN-based CSI feedback and recovery should consider 
the practical constraint of how often such feedback
can be transmitted and how CSI of
fading channels would vary due to the Doppler effect. 
\end{enumerate}

In order to reduce
computational complexity and model size, 
we seek to systematically exploit the physical channel characteristics 
such as the temporal coherence of forward CSI. 
Instead of training an RNN as a black box to learn and acquire the
underlying CSI characteristics for compression, feedback, and
recovery,  we directly leverage the known channel coherence 
temporally by developing a simple but effective Markovian model driven
differential CSI feedback framework MarkovNet to improve CSI recovery accuracy 
and reduce computational complexity. Spherical CSI feedback framework and enhanced CSI feedback network structure are proposed to provide the accurate prior information in the initial time slot. CNN-based CSI feedback networks are trained to further compress
and recover the differential CSI effectively. We show that this simple
MarkovNet can directly take advantage of the channel fading 
property to deliver much more efficient CSI compression and
recovery. 

This paper is organized as follows. Section~\ref{sec:system-mod} describes the massive MIMO system model commonly adopted in
this and similar works. Section~\ref{sec:coherence} presents two approaches to 
exploit CSI temporal coherence: RNNs and differential encoding. Section~\ref{sec:diff-net} describes our proposed differential encoding-based CSI feedback framework, MarkovNet, as well
as data pre-processing techniques to further improve CSI feedback accuracy for individual channel
snapshots such as power-based spherical normalization. Section~\ref{sec:model_cmp} introduces the proposed CNN-based dimension compression and decompression module for model size and complexity reduction.
Section~\ref{sec:evaluation} presents our experimental results, including computational analysis and performance under feedback quantization, for MarkovNet in comparison 
with a benchmark RNN-based network.
Section~\ref{sec:conclusion} concludes 
this manuscript. 

\section{System Model} \label{sec:system-mod}

\subsection{Forwardlink Channnel Estimation and Reconstruction}
In this paper, we consider a massive MIMO BS known in 5G as gNB equipped with $N_b \gg 1$ antennas
to serve a number of single-antenna UEs within its cell. 
We apply orthogonal frequency division multiplexing (OFDM) 
in downlink transmission
over $N_f$ subcarriers. 

To model the received signal of a UE,
consider the $m-$th subcarrier at time $t$.
Let $\mathbf{h}_{t,m} \in \mathbb{C}^{N_b\times1}$ denote the channel vector, 
$\mathbf{w}_{t,m} \in \mathbb{C}^{N_b\times1}$ denote transmit precoding vector, $x_{t,m}\in \mathbb{C}$ 
be the transmitted data symbol, and $n_{t,m}\in \mathbb{C}$ be the additive noise.
Then the received signal of the UE on the $m-$th subcarrier at time $t$ is given by
\begin{equation}
	y_{t,m} =\mathbf{h}_{t,m}^H\mathbf{w}_{t,m}x_{t,m} + n_{t,m}, 
\label{equ1}
\end{equation}
where $(\cdot)^H$ represents the conjugate transpose.
The downlink CSI matrix in the spatial frequency domain at time $t$ is denoted as $\tilde{\mathbf{H}}_t = \left[\mathbf{h}_{t,1},..., \mathbf{h}_{t,N_f}\right]^H \in \mathbb{C}^{N_f\times N_b}$.  

Based on the downlink channel matrix 
$\tilde{\mathbf{H}}_t $, the gNB can determine the transmit precoding vector for each subcarrier. 
However, since the size of the CSI matrix $\tilde{\mathbf{H}}_t$ is ${N_f}\times {N_b}$, 
the UE's CSI feedback payload is large and consumes a staggering amount of
uplink bandwidth in massive MIMO systems. 

\begin{figure}[!hbtp] \centering 
	\includegraphics[width=0.8\linewidth]{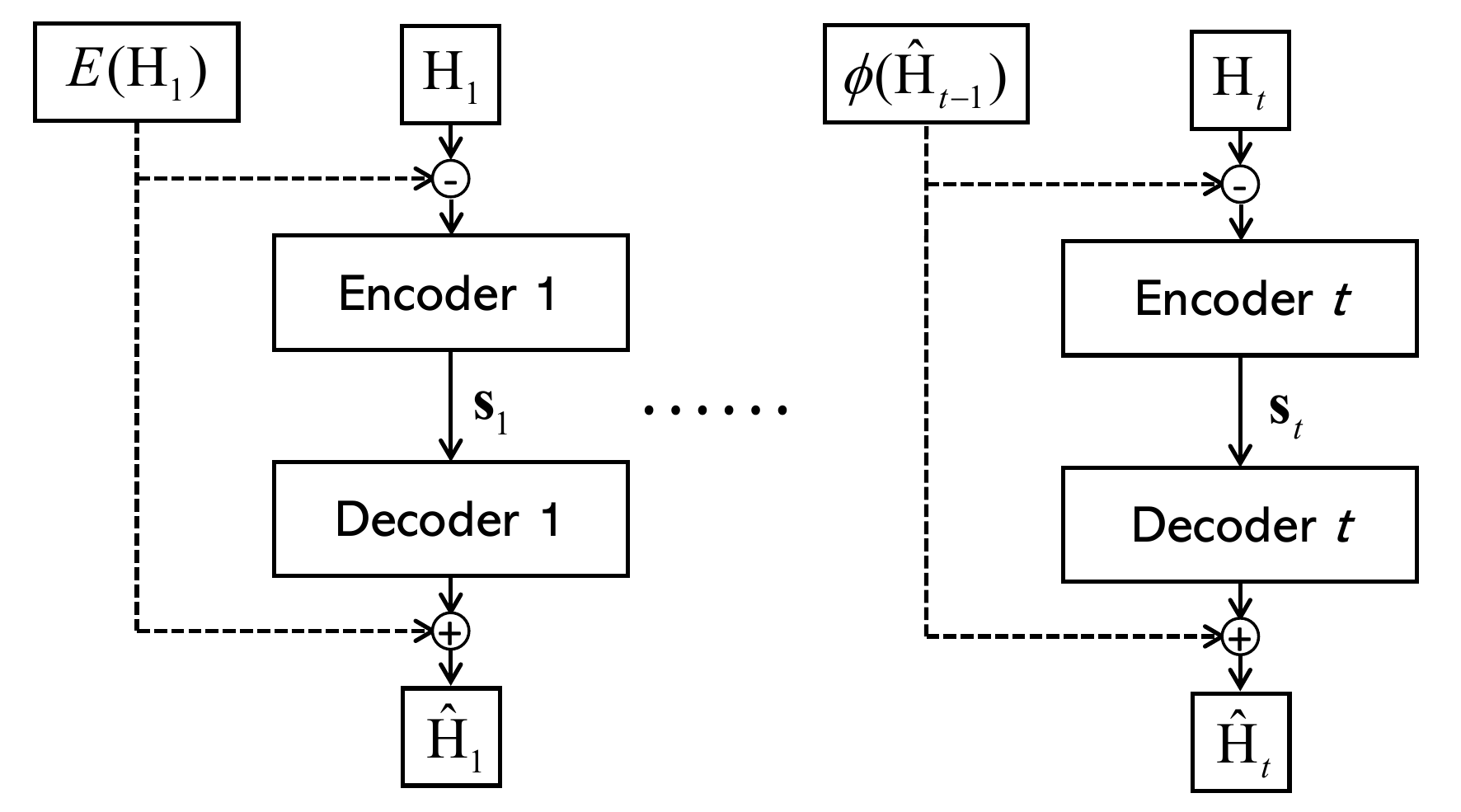}
	\caption{Illustration of the temporal correlation based CSI feedback. $(t>1)$} 
	\label{fig:function-model} \vspace*{-2mm}
\end{figure}

To reduce the feedback overhead, we first exploit the sparsity of CSI in a different projection space, the delay domain.
Multipath effects cause short delay spreads, 
resulting in sparse CSI matrices in the delay domain \cite{ref:sparsity}.
With the help of 2D discrete Fourier transform (DFT), CSI matrix 
$\mathbf{H}_f$ in 
spatial-frequency domain can be transformed  to be $\mathbf{H}_d$ in angular-delay domain using
\begin{equation}
	\mathbf{F}_d^H \mathbf{H}_f  \mathbf{F}_a =  \mathbf{H}_d, \label{idft3}
\end{equation}%
where $\mathbf{F_d}$ and $\mathbf{F_a} $ denote the $N_f \times N_f$ and $N_b \times N_b$ unitary DFT matrices, respectively.  
After 2D DFT of $\mathbf{H}_f$, most elements in the $N_f\times N_b$
matrix $\mathbf{H}_d$ are negligible except for the
first $R_d$ rows that dominate
the channel response\cite{ref:csinet}. Therefore, we can approximate the
channel by truncating CSI matrix to the first $R_d$ rows.  $\mathbf{H}_t$ is utilized to denote the first $R_d$ rows of matrices after 2D DFT of $\tilde{\mathbf{H}}_t$. 
Using $\mathbf{H}_t$ as a supervised learning objective, 
a DL based encoder and decoder,
which is often referred to as an autoencoder,
can be jointly trained
and optimized to achieve efficient 
CSI compression and reconstruction as shown in
Fig.~\ref{fig:function-model}(a). Several
recent works that adopted this autoencoder structure
\cite{ref:csinet}\cite{ref:dualnet} have
reported notable successes. 

To allow gNB to 
track the time-varying characteristics
of wireless fading channels, 
UEs need to periodically estimate and 
feed back instantaneous CSI with
high power and bandwidth efficiency. 
Considering a time duration with $T$ successive
time slots, the sequence of time-varying channel
matrix is defined as $\left\{\mathbf{H}_{t}\right\}_{t=1}^{T}=
\left\{\mathbf{H}_{1}, \mathbf{H}_{2}, \cdots, \mathbf{H}_{T}\right\}$. 

\subsection{High Efficiency CSI Feedback Encoding}
To reduce feedback overhead,
temporal coherence of the radio 
fading channels can be exploited. 
Since RF channels of mobile UEs are 
governed by physical scatters, multipaths,
bandwidth, and Doppler effect, 
the fading CSI exhibits physically coherent
characteristics including coherence time,
coherence bandwidth, and coherence space. 
For mobile users, coherence time 
measures temporal channel 
variations and describes the
Doppler effect caused by UE mobility. 
For most application scenarios, 
the massive MIMO channels do not vary abruptly.
By exploiting the channel coherence time, 
the UE and the gNB can rely on their previously
stored CSI estimates to encode only
the innovation components within the CSI.
Specifically, the UE can encode and feed back 
CSI variations instead of the full CSI to 
substantially reduce feedback cost. 
Accordingly, gNB can combine the new feedback with its
previously recovered CSI within coherence time 
to reconstruct subsequent CSI estimates. 

We can adopt a general first order Markovian  channel model
\begin{equation}
p(\mathbf{H}_t | \mathbf{H}_{t-1},\; \cdots,\; \mathbf{H}_1) = 
p(\mathbf{H}_t | \mathbf{H}_{t-1}).
\label{eq:1stMarkov}
\end{equation}
Given knowledge of the CSI at the previous time slot, 
the minimum mean square estimation (MMSE) of $\mathbf{H}_t $
can be defined as
\begin{equation}
\phi(\mathbf{H}_{t-1}) = E\{\mathbf{H}_t |\mathbf{H}_{t-1}\}.
\end{equation}
We define the MMSE estimation error as 
\begin{equation}
  \mathbf{V}_t=\mathbf{H}_{t}-E\{\mathbf{H}_t |\mathbf{H}_{t-1})
 = \mathbf{H}_{t}-\phi(\mathbf{H}_{t-1}).
\end{equation}
Consider the scenario that, at time $t-1$, 
the UE and the gNB have successfully
exchanged the CSI $\mathbf{H}_{t-1}$.
Then it would be more efficient for the UE
to compress and feed back the CSI estimation error
$\mathbf{V}_t$ to the gNB instead of the raw $\mathbf{H}_t$.

Based on this CSI model, we can develop a novel DL
encoder and decoder architecture by exploiting
a trainable neural network to learn the unknown 
MMSE estimation function $
\phi(\mathbf{H}_{t-1}) = E\{\mathbf{H}_t |\mathbf{H}_{t-1}\}.$
This new DL encoder and decoder architecture
is shown in 
Fig.~\ref{fig:function-model}. 

As shown in 
Fig.~\ref{fig:function-model}, the feedback for the CSI matrix sequence 
can be divided into two phases: a) The feedback of CSI at the first (initial) time slot ($t = 1$) without prior information; b) The feedback of CSI in subsequent time slots ($t = 2,3,...,T$) 
given the prior CSI information.
Denote $\mathbf{\hat{H}}_t$ as the reconstruction of  
CSI matrix $\mathbf{{H}}_t$ at time slot $t$. 
Define the encoding and decoding function as $f_{e}(\cdot)$ and $f_{d}(\cdot)$, respectively.
For downlink CSI feedback architecture in the
first time slot, the encoder network and decoder network can be denoted, respectively, by 
\begin{eqnarray}
\qquad\qquad	\mathbf{s}_1 &=& f_{e,1}(\mathbf{H}_{1}-E\{\mathbf{H}_1\}), 
\\
	\hat{\mathbf{H}}_{1} &=& f_{d,1}(\mathbf{s}_1) + E\{\mathbf{H}_1\}
\end{eqnarray}%

\begin{figure}[!hbtp] \centering 
	\includegraphics[width=0.95\linewidth]{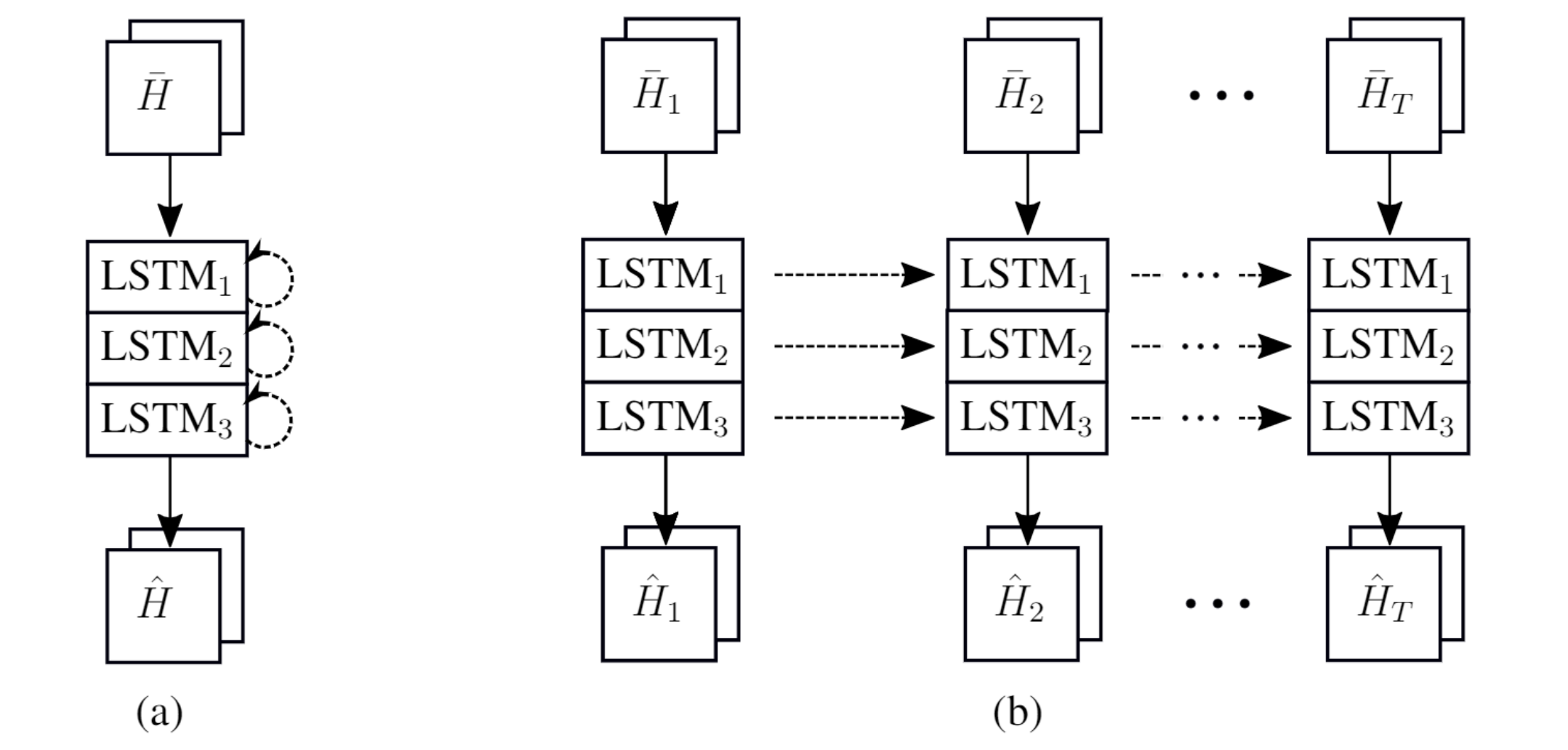}
    % \incfig{lstm-example}{0.95}
	\caption{(a) Illustration of a ``stacked'' LSTM network of depth 3 shown with recurrent connections. (b) Same network ``unrolled'' into $T$ timeslots. The network is trained with either perfect or quantized CSI, $\bar{ \mathbf{H}}$, to generate CSI estimates, $\hat{\mathbf{H}}$.} 
	\label{fig:lstm-example} \vspace*{-2mm}
\end{figure}

This initial phase assumes that the CSI mean is known 
from training or past information.  If such information is
unavailable, then $E\{\mathbf{H}_1\}=0$ shall be applied. 
For downlink CSI feedback architecture of subsequent
time slot $t$ ($t\ge 2$), the encoder network and decoder network can be executed, respectively, by 
\begin{eqnarray}
\qquad\qquad	\mathbf{s}_t &=& f_{e,t}(\mathbf{H}_{t}-\phi(\hat{\mathbf{H}}_{t-1})), 
\\
	\hat{\mathbf{H}}_{t} &=& f_{d,t}(\mathbf{s}_t)
	+\phi(\hat{\mathbf{H}}_{t-1})
	%\hat{\mathbf{H}}_{t-1}).
\end{eqnarray}%
Since the optimum function $\phi(\hat{\mathbf{H}}_{t-1}))$ is
unknown, one direct solution is to approximate the function
with deep neural network architecture trained by using a set
of CSI samples.

% Simplest = $g()$ is linear. Probably good literature support for this.

% \section{Experimental Evaluation} \label{sec:evaluation}

\section{Exploiting Channel Temporal Coherence} \label{sec:coherence}
We now discuss two avenues for exploiting the temporal coherence of
CSIs at successive time-slots: 
a DNN architecture that utilizes long-short term memory (LSTM) layers
and an information theoretic differential encoding approach. 

\subsection{Recurrent Neural Networks}
% -> prior work: rnns for csi estimation which motivates...
Recurrent neural networks (RNNs) include layers which encode memory of previous states. Through backpropagation 
training, recurrent layers learn whether to incorporate information stored in memory in the layer's output and whether that information should be kept in memory \cite{Hermans2013}. The memory 
incorporation enables RNNs to store, remember, and
process information that resides in
past signals for long time periods. RNNs can
utilize past input sequence samples to predict 
future states. \cite{Pascanu2014}.

RNNs have found wide applications 
in areas such as natural language processing (NLP), 
including machine translation \cite{ref:Sutskever2014} and sentiment extraction \cite{ref:Irsoy2014}. For NLP tasks,
empirical results have demonstrated 
the effectiveness of ``deep'' or ``stacked'' RNNs, networks which use the outputs of hidden recurrent layers as inputs to subsequent recurrent layers \cite{ref:Goldberg2016}. 

% Here we describe the application of RNN in connection
% with the RNN proposal of \cite{ref:csinet-lstm} for 
% massive MIMO CSI feedback. 
% Denote the hidden state of the $i$-th recurrent layer at the $t$-th timeslot as $\vec h_t^{(i)}$. A stacked RNN of depth $D$ takes,
% \begin{align}
%     \begin{split}
%         \vec h_t^{(i)} &= f\Big(\mathbf W^{(i)} \vec h_t^{(i)}+\mathbf W^{(i-1)} \vec h_t^{(i-1)}
%         % \\
%         % &\phantom{= f}
%      +\mathbf V^{(i)} \vec h_{t-1}^{(i)}+b^{(i)}\Big)
%     \end{split}
% \end{align}
% for $i>1$. For the first layer ($i=1$) with input vector, $x_t^{(i)}$, 
% \begin{align}
%     \vec h_t^{(1)} &= f\Big(\mathbf W^{(1)} \vec x_t^{(1)}+\mathbf V^{(1)} \vec h_{t-1}^{(1)}+b^{(1)}\Big)
% \end{align}
% with trainable parameters $\mathbf{W}^{(i)}, \mathbf{V}^{(i)}, $ and $b^{(i)}$ for all $i\in[1,\dots,D]$. The final output of the stacked LSTM network at timeslot $t$ is 
% \begin{align}
%     y_t &= g(\mathbf{U}\vec h_{t}^{(D)} +c)
% \end{align}
% with trainable parameters $\mathbf{U}, c.$

Prior works have investigated stacked RNNs for CSI estimation.
Several proposals have favored the 
use of Long Short Term Memory (LSTM) cell \cite{ref:csinet-lstm,ref:Lu2019,ref:Liao2019}, a recurrent unit that can tackle 
the vanishing gradient problem inherent in recurrent backpropagation \cite{ref:Hochreiter1997}. 
Existing LSTM-based works in CSI estimation have assumed that stacked LSTMs are better than shallow LSTMs, presenting models which used LSTM cells of depth 3 \cite{ref:csinet-lstm}. 
Fig.~\ref{fig:lstm-example} demonstrates the principle of
this LSTM network for CSI feedback and estimation. 
This bias towards deep RNNs is likely due to the aforementioned successes in NLP, where deep recurrent layers are theorized to learn hierarchical levels of semantic abstraction \cite{ref:Irsoy2014,ref:Bengio2009}. 

This RNN approach has been recently proposed in \cite{ref:csinet-lstm}.  In this work, we
shall consider the proposed architecture of 
\cite{ref:csinet-lstm} as the benchmark method. 
However, deep LSTMs can be problematic, as the number of parameters per LSTM cell can be quite large. If a parsimonious model is desired due to memory constraints, then memory intensive RNNs can be very costly. 

In this work, we explore ways to simplify the LSTM 
architectures without CSI performance loss, 
as deeper networks do not necessarily
lead to better estimation accuracy. 
In fact, we shall show later 
(Fig.~\ref{fig:lstm-vary}, Section~\ref{sec:rnn-for-csi})
that a single LSTM layer (i.e., $D=1$) could yield higher
accuracy CSI estimates than deeper LSTMs (i.e., $D\in[2,3]$) in some cases.

\subsection{CSI Entropy and Feedback Encoding}
% -> conditional entropy / efficient feedback / channel dependency
% -> motivates differential feedback in next section

% Despite initial successes and the heuristics for applying the
% complex RNN for CSI estimation and recovery, several
% important questions remain. 
Despite the success of deep RNNs in CSI estimation and recovery, 
several important research questions remain. 

\begin{itemize}
\item
First, what simplifications can be made to reduce
computational complexity while maintaining efficient CSI feedback 
and accurate CSI recovery?
\item
Second, how much CSI feedback bandwidth in terms of 
bitwidth per CSI coefficient is sufficient? 
\item
Third, how frequently should a UE should provide CSI feedback
for gNB to update its CSI estimate?
\end{itemize}
It is therefore important to tackle these open questions 
that hamper the
practical application and efficacy of DL based 
CSI estimation and recovery in massive MIMO networks. 

Consider random channel matrix $\mathbf{H}_t$
that consists of complex fading coefficients for
the $t$-th timeslot. We denote its joint probability density
function $p(\mathbf{H}_t)$ and define the corresponding 
entropy as
\begin{align}
    H(\mathbf{H}_t) &= -\sum_{\mathbf{H}_t} p(\mathbf{H}_t)\log p\left(\mathbf{H}_t\right) \label{eq:entropy}
\end{align}
where (\ref{eq:entropy}) is the sum over all realizations of r.v. $\mathbf{H}_t$. 
The CSI entropy of (\ref{eq:cond-joint}) describes the
required number of bits for the UE to feed back its
CSI estimate to the gNB for reconstruction. 
Denote the $(i,j)$-th CSI element within $\mathbf{H}_{t}$
as $\mathbf{H}_{t, (i,j)}$ 
at time $t$. If all elements are independent,
then we have a simple upper bound on the entropy of the full CSI matrix as
\begin{align}
    H(\mathbf{H}_{t}) \leq H_{\rm UB}=\sum_{i,j} H(\mathbf{H}_{t,(i,j)}) \label{eq:joint-bound}
\end{align}
This entropy bound $H_{\rm UB}$ describes the approximate number
of total bits necessary for direct encoding
of forwardlink CSI for UE feedback. 

\begin{figure}[thpb]
	\centering
	\includegraphics[scale=0.9]{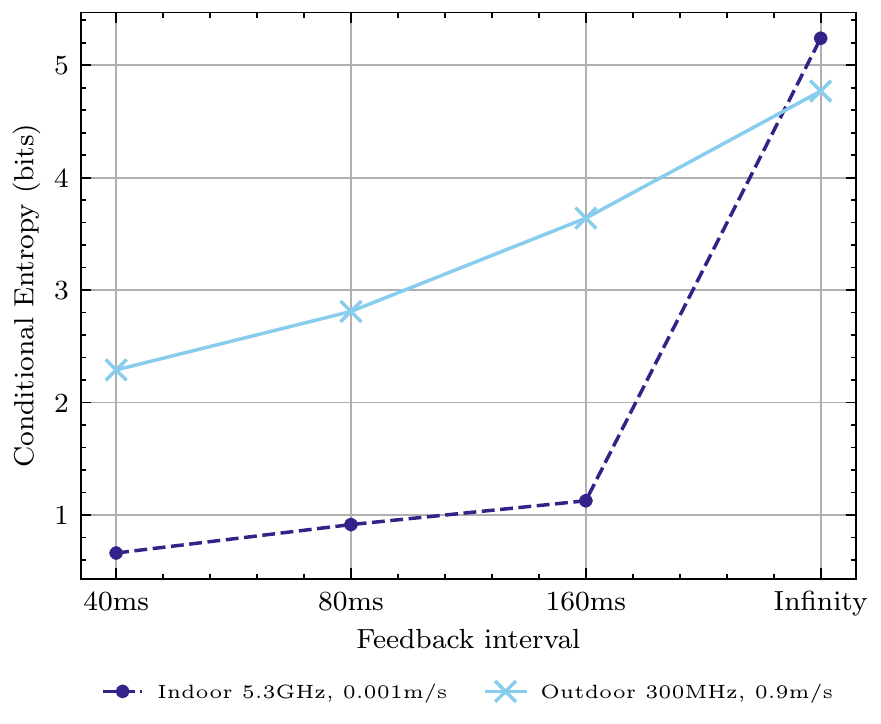}
% 	\caption{Conditional entropy, $H(\mathbf{H}_{t}|\mathbf{H}_{t-\delta})$ (bits/element), under different feedback intervals ($\delta$). }
	\caption{ Averaged conditional entropy (bits/element), under different feedback intervals ($\delta$). }
	\label{fig:conditional_entropy}
  %   \vspace*{-3mm}
\end{figure}

Fortunately, in mobile wireless networks, 
CSI within a coherence time exhibits strong correlation
\cite{ref:tse2005fundamentals}.
Therefore, instead of constructing CSI independently
by relying on CSI feedbacks for individual time slots, 
the gNB can utilize this CSI dependency by leveraging
both previously reconstructed CSIs and the current CSI feedback.
In other words, the UE feedback should focus on providing
information that is not available at the gNB from 
CSIs of previous time slots. 

Taking advantage of the Markovian  CSI model, we can investigate
how much the gNB can benefit from the previous CSI. 
Given the Markovian  channel model of (\ref{eq:1stMarkov}), 
the conditional CSI entropy quantifies the amount of information needed to characterize the CSI matrix based on the
available CSIs from earlier reconstruction:
\begin{align}
H(\mathbf{H}_{t}|\mathbf{H}_{t-1},\dots, 
\mathbf{H}_1)  &= H(\mathbf{H}_{t}|\mathbf{H}_{t-1}) 
\label{eq:cond-joint}\\
    &= -\sum_{\mathbf{H}_{t-1}}\sum_{\mathbf{H}_{t}}p(\mathbf{H}_{t})\log p(\mathbf{H}_{t}|\mathbf{H}_{t-1}) \nonumber
%    &= H(\mathbf{H}_{t},\mathbf{H}_{t-1}) - H(\mathbf{H}_{t-1}) 
\end{align}
From the well known relationship of $H(\mathbf{H}_{t}|\mathbf{H}_{t-1})\le H(\mathbf{H}_{t})$,
it is clear that by utilizing the most recently reconstructed
CSI, the gNB would require less feedback bandwidth
and improve the UE feedback efficiency. 

A stationary first order Markovian  channel model is
characterized by the conditional probability density
function of $p(\mathbf{H}_t|\mathbf{H}_{t-1})$.
In practice, such
distribution information on CSI is difficult to acquire
analytically. 
To gain valuable insights into the time-coherence
between CSI at different feedback intervals, 
we shall provide a numerical evaluation of typical
wireless channel models by comparing the entropy
and the conditional entropy of the forwardlink CSI
parameters. 
Note the following relationship between CSI entropies
at $t$ and $t-\delta$ where $\delta$ is the feedback
interval:
\begin{align}
    H(\mathbf{H}_{t, (i,j)}|\mathbf{H}_{t-\delta}) \le
    H(\mathbf{H}_{t, (i,j)}|\mathbf{H}_{t-\delta,(i,j)}) &
\leq H(\mathbf{H}_{t,(i,j)}). \label{eq:cond}
\end{align}
For practical reasons, 
we shall numerically evaluate the conditional entropy
of $  H(\mathbf{H}_{t, (i,j)}|\mathbf{H}_{t-\delta,(i,j)})$
averaged over the coefficients in $\mathbf{H}_t$.
Such information can present important guidelines to
the determining number of bits for CSI feedback
and how often UE should provide such CSI feedback 
for the CSI estimation of massive MIMO 
systems by the gNB. 

In this experiment, 
we consider the link with $N_b=32$ transmit antennas
and 1 receive antenna over $N_f=1024$ subcarriers.
After applying the 2D DFT, 
$R_d=32$ rows of significant CSI elements in delay domain
are retained in $\mathbf{H}_t$. 
For each element within $\mathbf{H}_t$, we
 apply a 14-bit uniform quantizer
to encode raw CSI values, resulting in
a normalized mean square quantization error of -40dB. Since the complex CSI matrices were always divided into real part and imaginary part as the  real-valued input to the neural network \cite{ref:csinet,ref:csinet-lstm,ref:Lu2019,ref:Liao2019,ref:Li2020,ref:Jang2019}, we consider the conditional entropy of the CSI's real part and imaginary part individually.
Fig.~\ref{fig:conditional_entropy} demonstrates the estimated
conditional entropy averaged over the $2 \times N_b\times R_d$
CSI elements.

We generate 10,000 random indoor and outdoor channel
responses using the channel models given in \cite{ref:cost2100} and 
\cite{ref:csinet-lstm}. 
Following the examples in \cite{ref:csinet-lstm},
the indoor channel is in the 5.3 GHz band, with little
or no mobility at velocity of 10$^{-3} $ m/s. 
The outdoor channel is in the 300 MHz band, at
velocity of 0.9 m/s. The bandwidth for indoor and outdoor channels is 20 MHz. The conditional entropy
is evaluated for different lengths
of feedback interval $\delta=40$ms, 80ms, 160ms, and
$\infty$ (i.e., no feedback).

As the results of Fig.~\ref{fig:conditional_entropy}, the
average conditional entropy varies between 1 to 5 bits/element
for both the indoor and outdoor channel models tested. 
As the duration of the feedback interval grows, the conditional 
entropy increases because of the limited channel coherence time.
In addition, it is intuitive that the outdoor channel 
exhibits higher conditional entropy since higher velocity
corresponds to shorter coherence time \cite{ref:rappaport1996wireless}.
For both channel models, the average
entropy of the CSI elements without prior CSI 
can be seen for $\delta=\infty$ which attends its maximum
value of approximately 5 bits. These numerical
results strongly motivate a systematic selection of feedback
interval and feedback bandwidth. For example,
for a feedback interval of 80 ms, an average of
approximately 3 bits per CSI coefficient
can be used for the outdoor CSI feedback by UEs when
prior CSI is utilized by the gNB. On the other hand,
for the same feedback interval, an average bitwidth of
1 bit per CSI coefficient can be used for indoor CSI feedback. 

The reduction of CSI entropy under conditions of
known prior CSI knowledge motivates the idea of
condition-based encoding such as
differential encoding by the UE. 
Encoding the difference between 
successive feedback instants,
$\mathbf{H}_t$ and $\hat{\mathbf{H}}_{t-\delta}$,
can reduce the required number of UE feedback bits,
allowing more compression without loss of information \cite{ref:Dhanani2013}.

\section{Differential CSI Encoding} \label{sec:diff-net}
\subsection{A Simplified Markovian  Model}
Although the general Markovian  CSI model motivates 
the use of a DNN
to approximate the conditional mean
$ \phi(\hat{\mathbf{H}}_{t-1})$ through training
and learning, we can examine 
simpler CSI models in order to develop 
a low complexity encoder-decoder
structure with consistently strong performance.
Consider the simplified
Markovian  CSI model of \cite{huber2006improved}:
\begin{equation}
    \mathbf{H}_{t} = \gamma \mathbf{H}_{t-1} + \mathbf{V}_t
    \label{eq:ARmodel}
\end{equation}

\begin{figure}[thpb]
	\centering
	\includegraphics[width=\columnwidth]{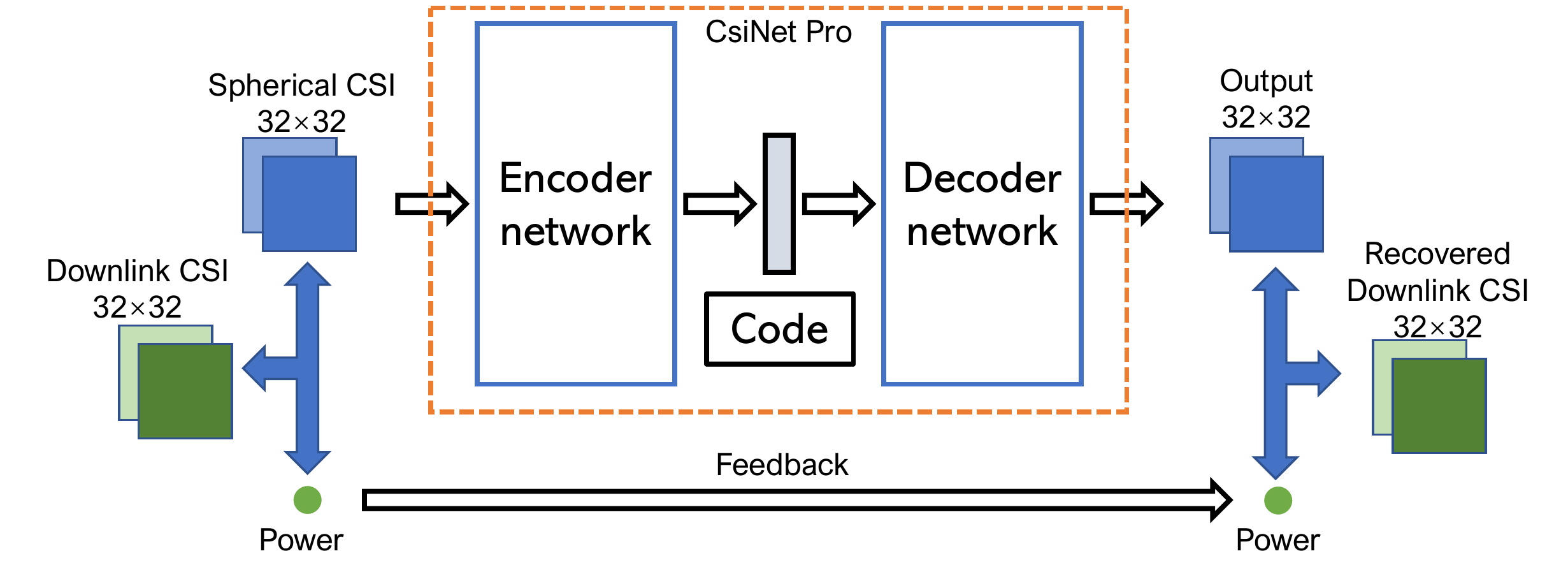}
	\caption{Architecture of spherical CSI feedback in SphNet.}
	\label{figure_sph}
  %   \vspace*{-3mm}
\end{figure} 

where $\gamma$ is a constant and $\mathbf{V}_t$ is a zero-mean
i.i.d. random matrix. Given ensemble samples of $\mathbf{H}_{t-1}$
and $\mathbf{H}_{t}$, the unknown $\gamma$ can be estimated via
\begin{equation}
  \hat{\gamma} = \frac{ \mbox{Trace}(E\{\mathbf{H}_{t}\mathbf{H}_{t-1}^H  \})}
   {E\| \mathbf{H}_{t-1} \|^2 }.
\end{equation}

Based on this simplified 1st order autoregressive (AR) model,
we propose a low complexity encoder-decoder architecture
for time slot $t$ ($t\ge 2$) as
\begin{eqnarray}
\qquad	\mathbf{s}_t &=& f_{e,t}(\mathbf{H}_{t}
-\gamma \hat{\mathbf{H}}_{t-1}), \label{eq:diffinfo}
\\
	\hat{\mathbf{H}}_{t} &=& f_{d,t}(\mathbf{s}_t)
	+\gamma \hat{\mathbf{H}}_{t-1}\label{eq:diffdec}
	%\hat{\mathbf{H}}_{t-1}).
\end{eqnarray}%

Based on this simplified model, 
we propose a differential encoding 
architecture named ``MarkovNet''
for efficient CSI feedback and reconstruction
in the massive MIMO systems. The 
differential CSI feedback framework consists of 
two phases: a first network used for the initial CSI at $t_1$ (Spherical CSI feedback), and a second network in subsequent 
timeslots to compress 
and encode differential information as described by
the encoder of
(\ref{eq:diffinfo}).

% \begin{table}[]
% \centering
% \renewcommand{\arraystretch}{1.5}
% \caption{Conditional entropy, $H(\mathbf{H}_{t}|\mathbf{H}_{t-t_0})$, (bits/element) under different speeds and feedback intervals ($t_0$). The baseline corresponds to the entropy without prior information, $H(\mathbf{H}_{t})$.}
% \begin{tabular}{|c|c|c|c|c|c|}
% \hline
%                         & \textbf{baseline} & \textbf{40ms} & \textbf{80ms} & \textbf{160ms} & \textbf{320ms} \\ \hline
% \textbf{outdoor 0.9m/s} & 4.77              & 2.29          & 2.81          & 3.64           & 4.41           \\ \hline
%                         & \textbf{baseline} & \textbf{5ms}  & \textbf{10ms} & \textbf{20ms}  & \textbf{40ms}  \\ \hline
% \textbf{outdoor 3.6m/s} & 4.61              & 1.53          & 2.03          & 2.65           & 3.42           \\ \hline
% \end{tabular}
% \label{tab:conditional_entropy}
% \end{table}

To fully exploit the temporal CSI coherence, 
accurate CSI at the initial time slot 
$t_1$ is required to provide sufficient 
baseline information for the CSI feedback in 
subsequent time slots. 
To this end, our proposed framework
shall apply CSI pre-processing and 
optimize the neural network structure. Specifically,
\begin{itemize}
    \item We propose the spherical normalization
    for CSI pre-processing to the input data distribution to make the network's objective function more applicable to the commonly adopted accuracy metric, NMSE.
    \item We also optimize the CSI encoder-decoder to enhance CSI recovery accuracy.
\end{itemize}

\subsection{Transforming CSI Feedback in Spherical Coordinate}

How to effectively apply DL techniques to exploit channel data properties and optimization objects remains an open research issue, 
as
many existing DL based works mainly utilize the deep learning architectures and optimization functions 
successfully developed for other application areas.
Direct adoption of DL architectures
without customization for CSI data characteristics risks 
unsatisfactory performance.
In particular, data processing methods and loss functions 
developed for computer vision may not be
well suited for CSI encoding and reconstruction. 

To begin, many 
existing DL-based CSI encoding-decoding schemes
conveniently view the 2D MIMO channel matrix $\mathbf{H}_t$ 
as akin to an image such that the 
normalized elements of 
the CSI matrix are utilized as image-like
input to DL networks in both training and testing. 
However, the multipath fading MIMO channels exhibit
unique special properties and probability distributions
different from 2D image data.

Among other differences,
images are represented as matrices of intensity 
pixel values. 
For color images, each color channel
corresponds to a 2D matrix 
of pixel values that are unsigned integers, e.g.,
in the range between 0 and 255. 
By normalizing these pixels, there can be strong benefit in preparing the images as inputs of the DL model. 
However, unlike different images whose pixel values
are mostly in the same order of magnitude, 
the dynamic range of CSI data can be much greater.
For example, RF pathloss grows polynomially with distance 
between gNB and UE\cite{ref:wireless}. 
As a result, CSI of one UE can be different from
CSI of another UE by several orders of magnitude,
depending on their respective distances to gNB. 
A naive normalization can render CSIs of some
UEs too small for the DL networks to respond to. 
Another different feature is that the baseband
CSI parameters are complex values, consisting of
both magnitude and phase, whereas image pixels are
nonnegative real (with normalization).

In terms of learning objectives, several
existing DL-based CSI feedback works adopted the 
loss function similar to image recovery for training
the DL networks. Specifically, the objective is to minimize
the mean square error (MSE):
\begin{equation}
	\textrm{MSE} = \frac{1}{N}\sum_{k=1}^N
	\Arrowvert\mathbf{H}_{k}-\mathbf{\hat{H}}_{k}\Arrowvert^2,
\end{equation} where $k$ and $N$ are, respectively, the index and total number of samples in the data set,
whereas $\left \|  \cdot\right \|$ denotes 
the Frobenius norm. On the other hand, it is typical more
meaningful in CSI estimation to apply
the normalized MSE (NMSE)
\begin{equation}
	\textrm{NMSE} = \frac{1}{N}\sum_{k=1}^N\Arrowvert\mathbf{H}_{k}-\mathbf{\hat{H}}_{k}\Arrowvert^2/\Arrowvert\mathbf{H}_{k}\Arrowvert^2,
	\label{NMSE}
\end{equation}
to assess the accuracy of CSI recovery
at the gNB \cite{ref:channel_estimation} and feedback efficiency \cite{ref:csinet,ref:dualnet,ref:csinet-lstm}. 
By directly using MSE as the loss function,
the DL networks would be biased toward
the CSI accuracy of stronger MIMO channels.

In response to the domain-specific characteristics of  data
and objective in CSI estimation, we propose to use a spherical CSI 
data structure for feedback as shown in Fig. \ref{figure_sph}. The spherical CSI feedback architecture splits the downlink CSI matrix $\mathbf{H}_{k}$ into a magnitude value $p_k$ and a 
spherical downlink CSI matrix $\check{\mathbf{H}}_{k}$, where 
$p_k = \Arrowvert\mathbf{H}_{k}\Arrowvert$ is the size
of the CSI matrix whereas the unit norm 
spherical CSI $\check{\mathbf{H}}_{k} = \mathbf{H}_{k}/\Arrowvert\mathbf{H}_{k}\Arrowvert$ represents
remains on the surface of the unit hyper-sphere.
The UE would encode and feedback both
the size $p_k$ and the 
spherical CSI matrix $\check{\mathbf{H}}_{k}$ separately.

\begin{figure}[thpb]
	\centering
	\includegraphics[scale=0.5]{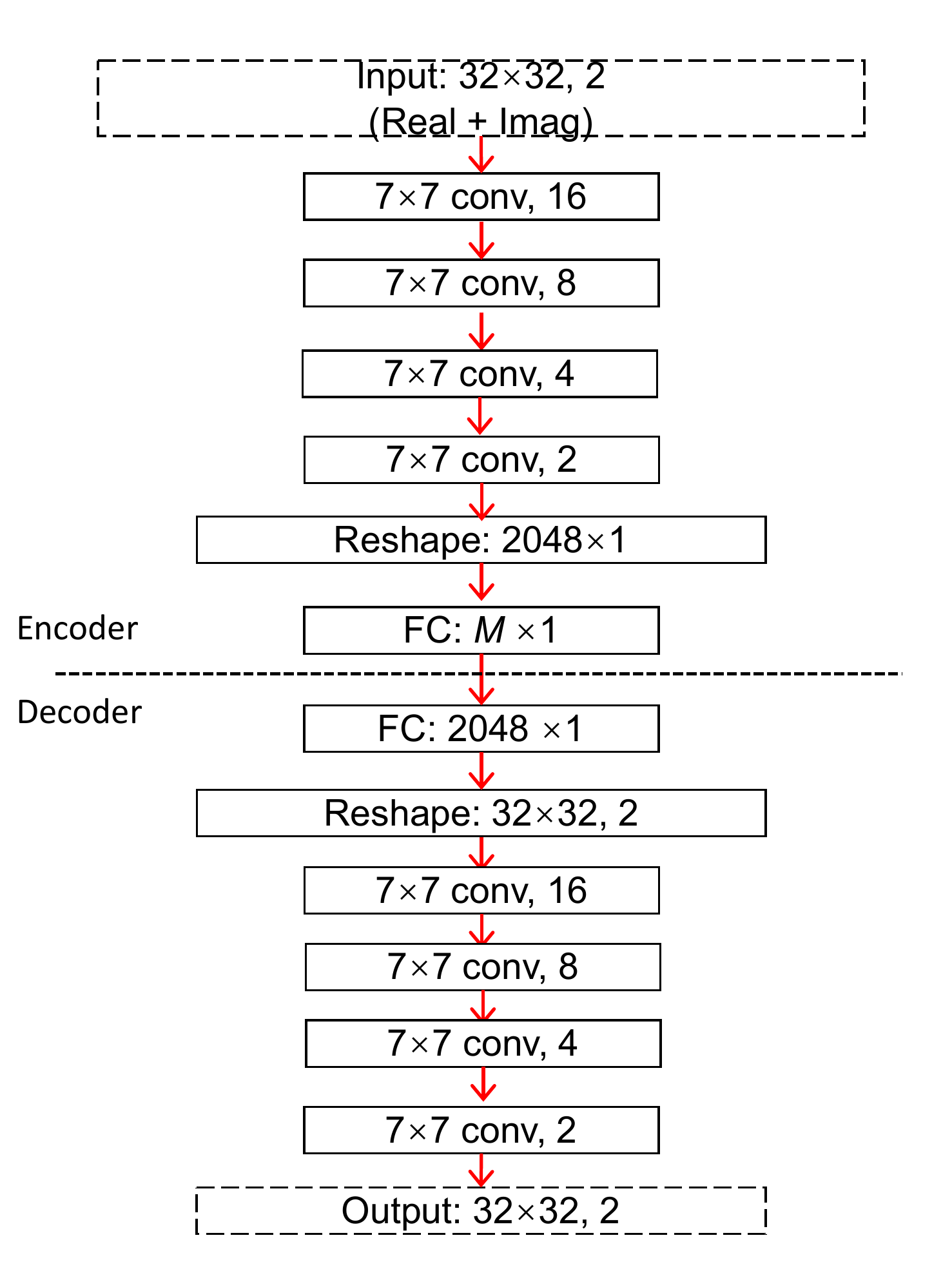}
	\caption{Architecture of CsiNet Pro.}
	\label{fig:csinet_pro}
  %   \vspace*{-3mm}
\end{figure}

Spherical CSI feedback architecture presents
numerical advantages. First,
we can construct an encoder DL network that focuses on
compressing and encoding the spherical CSI matrix $\check{\mathbf{H}}_{k}$. The size of the CSI would
be directly sent back to the gNB separately since 
it contains little redundancy. Thus, even for CSI matrices 
of different magnitudes, they are equally important in training 
the encoder and decoder networks.
During training the gradients for UEs that are far away from
the gNB would no longer
be negligible \cite{ref:normalization}. Moreover, 
 the decoder will have 
a more limited domain for more accurate CSI recovery under spherical normalization \cite{Liu2020SphNet}.

As shown in Fig.~\ref{figure_sph}, our joint
CSI compression and reconstruction architecture
still utilizes the effective autoencoder 
structure in which the encoder at the
UE attempts to learn a low-dimensional 
CSI representation for a relatively high-dimensional 
dataset represented in the form of spherical CSI matrices. 
The decoder at the gNB reconstructs the CSI matrix
based on feedback information extracted from the UE encoder 
and the direct feedback of CSI magnitude $p_k$. 

\subsection{CsiNet Pro: A Novel CSI Encoder-Decoder Network}

We propose an efficient neural network structure, named CsiNet Pro,
for UE encoding and gNB decoding of CSI in massive MIMO
networks. The structure of CsiNet Pro is illustrated
in Fig.~\ref{fig:csinet_pro}. 
In comparison with existing neural networks such as
those from \cite{ref:csinet}\cite{ref:csinet-lstm}, CsiNet Pro 
provides a deeper encoder that uses
more convolutional layers to better extract features of CSI.
There is a corresponding decoder at the gNB that also
contains 4 convolution layers. 

The design of encoder for dimension compression is crucial. 
However, the encoders in \cite{ref:csinet,ref:dualnet, ref:csinet-lstm} all utilized one convolutional layer and one fully connected layer. As a major departure, the encoder
of CsiNet Pro utilizes 4 convolutional layers for feature extraction and 1 fully connected layer for dimension compression. 
Specifically, the 4 convolutional layers
apply $7\times 7$ kernels to generate $16$, $8$, $4$ and $2$ feature maps, respectively (see Fig.~\ref{fig:csinet_pro}). 

\begin{figure*}[!hbtp] \centering 
	\includegraphics[width=0.8\linewidth]{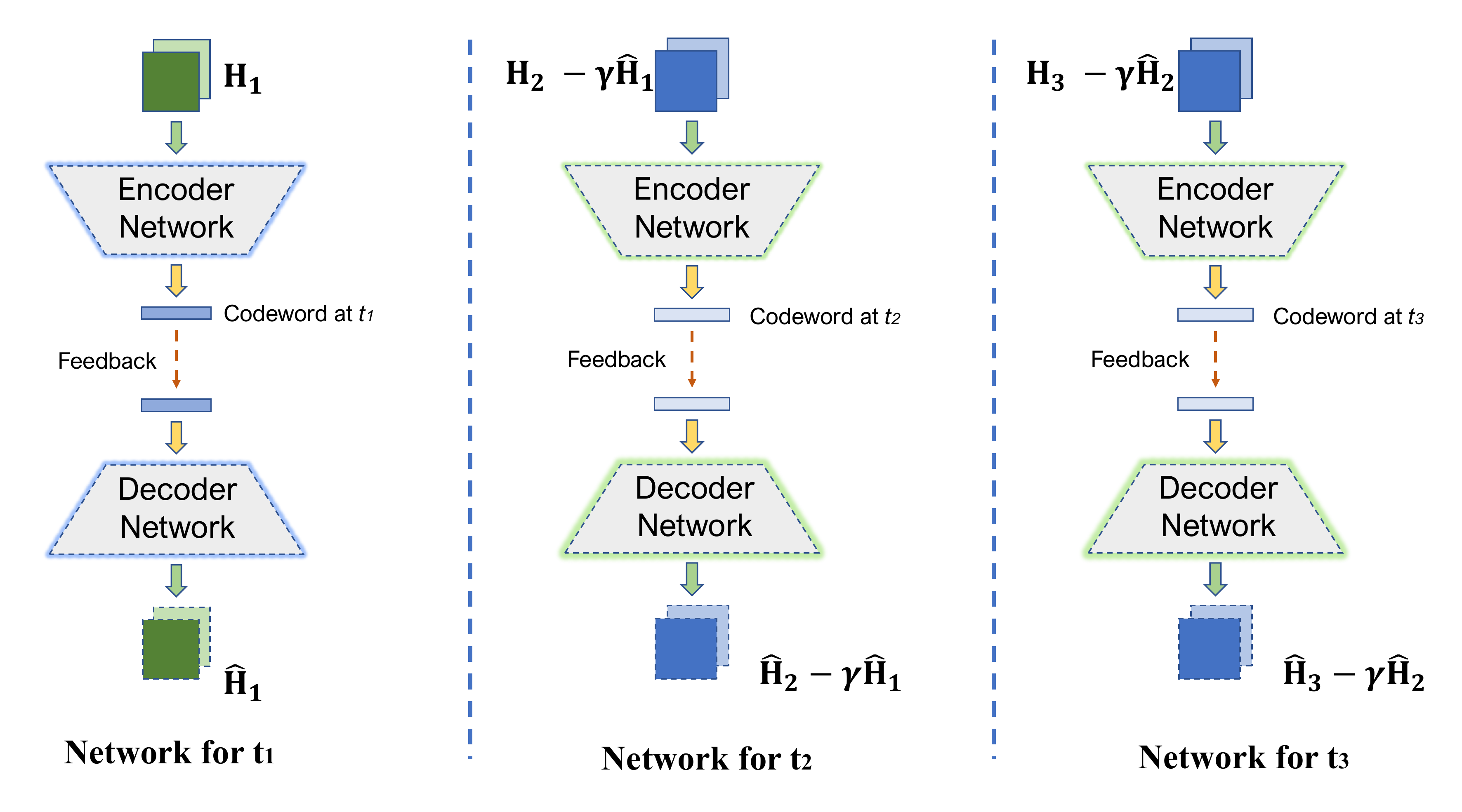}
	\caption{Illustration of the 
	multi-stage, differential CSI feedback framework MarkovNet.} 
	\label{fig:diffnet} \vspace*{-2mm}
\end{figure*}

Another major change in CsiNet Pro 
is the use of a different normalization range 
and output activation function. 
Recall that the decoder network utilizes 4 convolutional layers 
as shown in Fig.~\ref{fig:csinet_pro}. 
Unlike the nonnegative pixel values in image reconstruction,
CSI values contain both real and imaginary parts that
can be either positive or negative. 
Thus, unlike previous works that normalize the CSI values to 
fall within $[0, 1]$ in order to use “sigmoid” or “ReLU” as the activation function of the last
layer, our proposed CsiNet Pro normalizes the real
and imaginary CSI values to the range [-1, 1]
while using ``tanh" as its activation function 
in the last layer. 

We integrate the CsiNet Pro as part of
the spherical CSI feedback framework shown in 
Fig.~\ref{figure_sph} to enhance the CSI recovery accuracy.

\subsection{Differential CSI Encoding}

Motivated by the simplified first order
AR model for CSI, we
propose a differential CSI feedback framework MarkovNet
to improve bandwidth efficiency. 
Different from the RNN based networks such as LSTM
which relies on neural networks to learn the 
required information sharing and corresponding 
CSI compression simultaneously, MarkovNet 
proactively leverages the simplified AR model
(\ref{eq:ARmodel}) for CSI and encode the CSI
prediction error as shown in (\ref{eq:diffinfo})
between two successive time slots.

Recall that the difference 
based on first order estimation
of the CSI in two adjacent time slots $\mathbf{H}_t-\hat \gamma \mathbf{H}_{t-1}$ is an approximation of the innovation $\mathbf{V}_t$.
As shown in Fig. \ref{fig:diffnet}, for time-slots beyond the
initial time-slot, the linear prediction difference
$\mathbf{H}_t-\hat \gamma \mathbf{H}_{t-1}$ is sent to
the encoder network to execute the
encoding process of $ \mathbf{s}_t = f_{e,t}(\mathbf{H}_{t}
-\gamma \hat{\mathbf{H}}_{t-1})$ given in (\ref{eq:diffinfo}).
At the 
gNB receiver, the decoder network
can utilize the previously recovered
CSI $\hat{\mathbf{H}}_{t-1}$ to reconstruct $\hat{\mathbf{H}}_t$
according to $\hat{\mathbf{H}}_{t} = f_{d,t}(\mathbf{s}_t)
+\gamma \hat{\mathbf{H}}_{t-1}$ as described in (\ref{eq:diffdec}).

MarkovNet from $t_2$ onward would employ the same
network architecture CsiNet Pro as shown in 
Fig. \ref{fig:csinet_pro}. Compared with network for $t_1$ which uses a 
larger compression ratio to ensure the high recovery accuracy
in the first timeslot, MarkovNet from $t_2$ can 
afford smaller compression ratio to achieve a higher 
bandwidth efficiency with the help of prior information.

MarkovNet exhibits several additional advantages in
practical implementation. First, 
compared to RNN-based CSI feedback, 
MarkovNet can exploit pretrained model as initial 
neural network parameters for models used in later timeslots to improve training efficiency since the CSI at adjacent time slots share similar data features. 
Second, differential CSI matrix 
tend to be more sparse, hereby enabling MarkovNet to achieve 
a higher degree of compression during feedback. Third,
for most wireless network applications, both gNB and UEs have
limited power, computation, and storage resources. 
MarkovNet simplifies the learning tasks of neural networks 
and is more applicable in a wider variety of
wireless deployment scenarios. 

% Moreover,
% since the CSI feedbacks at adjacent time slots share similar data features, the pre-trained model in earlier time slots of MarkovNet 
% can be used as a good initialization for subsequent time slots
% to improve training efficiency and accuracy. 

% \begin{equation}
%     rank(\mathbf{H}_t-\mathbf{H}_{t-1}) \leq rank(\mathbf{H}_t)
% \end{equation}
% or $l_0$ norm

% \begin{itemize}
%     \item The difference is calculated by hand, and compressed by the neural network. Design a simple method/network for the decoder part
% \end{itemize}

\section{Model Reduction} \label{sec:model_cmp}

Practical implementation of deep
neural networks for CSI feedback and recovery
can be challenging to some mobile devices. 
Because DL network architectures often use 
large numbers of parameters,
they require substantial computation and memory resources.
Unrolled RNN models, such as the LSTM layers 
in Fig.~\ref{fig:lstm-example} 
are particularly computationally expensive.
For example, CsiNet-LSTM \cite{ref:csinet-lstm} at
a compression ratio (CR) of $1/16$ contains $1.19 \times 10^8$ parameters per timeslot. % 118,493,816
One of the main advantages of MarkovNet (see Fig.~\ref{fig:diffnet})
is its relatively low parameter count, as
a comparable version of MarkovNet at a CR of $1/16$ has $5.43\times 10^{5}$
parameters per timeslot, a difference of three orders of magnitude relative to CsiNet-LSTM.

\begin{figure*}[!hbtp] \centering 
	\includegraphics[width=0.9\linewidth]{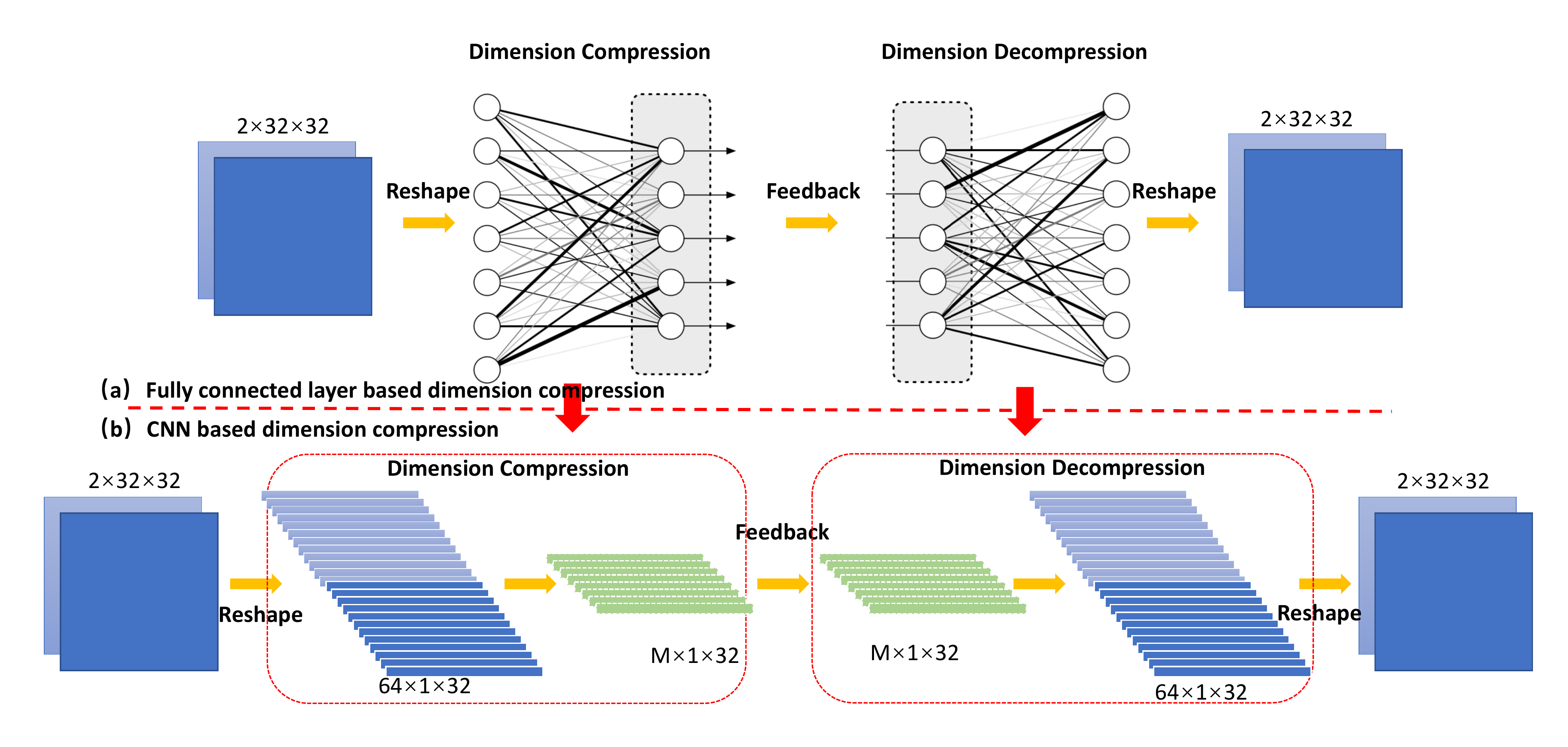}
	\caption{Proposed CNN-based dimension compression module.} 
\label{fig:cnn_compression} \vspace*{-2mm}
\end{figure*}

% To reduce the DL model size, we utilize the principle of
% differential CSI feedback based on the first order
% AR model.
Our proposed MarkovNet can clearly reduce the model size by eliminating the repeated structure 
used to learn the from the sequence data in RNN-style
architecture. 
It is important to note, however, the fully connected 
(FC) layers
for dimension compression and decompression 
in the current MarkovNet still contains
a large number of parameters. For example, there are
more than $10^{6}$ parameters for the FC layers at CR = $1/8$. 

The FC layers for dimension compression and decompression,
as shown in Fig. \ref{fig:cnn_compression}(a),
has often been adopted in deep learning based CSI feedback \cite{ref:csinet,ref:csinet-lstm,ref:dualnet,ref:Liao2019,ref:Lu2019}.
However,  elements of the CSI matrix 
only exhibit strong correlation with its neighbors 
in angular-delay domain. Thus, we recognize that the
FC layers, though effective and popular, 
still contains a large fraction of 
non-essential connections with
very weak weight parameters. This realization presents 
another opportunity for model reduction. 
To further reduce model size, we propose a 
CNN-based latent structure to replace the FC layers
for dimension compression. As shown in 
Fig. \ref{fig:cnn_compression}(b), we slice the 
two square feature maps into $64$ feature maps 
of dimension $1 \times 32$. We then design
$M$ CNN kernels of length $1 \times 7$ to compress the codewords dimension. The integer $M$ is adaptive in accordance
with the encoder compression ratio denoted by $\frac{M}{64}$. 
Through this feature processing, connections between
CSI elements that are far apart in the angular-delay domain 
are removed. Strongly correlated features of CSI matrix 
across the angular-delay domain can effectively 
be captured by the small CNN kernels.

\begin{table}[htb]
\centering
\renewcommand{\arraystretch}{1.5}
\caption{Number of parameters and FLOPs comparison for FC-based and proposed CNN-based dimension compression and decompression module. M: million, K: thousand.}
\label{tab: param-flops}
\begin{tabular}{|c|c|c|c|c|}
\hline
\multirow{2}{*}{\textbf{}}  & \multicolumn{2}{c|}{\textbf{Number of parameters}} & \multicolumn{2}{c|}{\textbf{FLOPs}}   \\ \cline{2-5} 
                            & \textbf{FC-based}        & \textbf{Proposed}       & \textbf{FC-based} & \textbf{Proposed} \\ \hline
\textbf{CR=$1/4$}  & 2.1 M                & 14.4 K                  & 4.2 M         & 0.9 M           \\ \hline
\textbf{CR=$1/8$}  & 1.1 M                & 7.2 K                   & 2.1 M         & 0.5 M           \\ \hline
\textbf{CR=$1/16$} & 0.5 M                  & 3.7 K                   & 1.0 M         & 0.2 M           \\ \hline
\end{tabular}
\end{table}

To illustrate the effect of the proposed model size reduction, 
we summarize the number of parameters and the floating point 
operations (FLOPs) in Table \ref{tab: param-flops}.
This information provides a comparison of the storage size
and computational complexity between the use
of FC-layer and proposed CNN-layer in CSI
compression module and the corresponding decompression module. 
As shown in Table \ref{tab: param-flops}, 
the proposed CNN-based dimension compression and decompression module reduces the 
number of parameters by over 100 times and
the number of FLOPs by at least 4 times. 
The comparison results demonstrate that our new 
CNN design for CSI compression and decompression 
represents an important step in broadening the 
range of practical applications for effectively
deploying deep learning based CSI encoding, 
feedback, and reconstruction in massive MIMO wireless systems. 

% \subsection{Feedback Interval}
% \begin{itemize}
%     \item A discussion about the feedback interval used in the standards and the way to adjust the feedback interval?  
%     \item By adjusting the feedback interval, our model can handle the CSI feedback in high mobility cases.
% \end{itemize}

\section{Performance Evaluation} \label{sec:evaluation}

We assess the performance of both RNN-based CsiNet-LSTM \cite{ref:csinet-lstm} and MarkovNet for two different massive MIMO scenarios generated from the well known
COST 2100 model \cite{ref:cost2100}.
\begin{enumerate}
\item \textbf{Indoor} channels using a 5.3GHz downlink at
0.001 m/s UE velocity, served by a
gNB at center of a $20$m$\times 20$m coverage area.
\item \textbf{Outdoor} channels using a 300MHz downlink at 0.9 m/s UE velocity served by a gNB at center 
of a $400$m$\times 400$m coverage area.
\end{enumerate}

We give $N_b=32$ antennas to the gNB to
serve single antenna UEs randomly distributed within
the coverage area. 
We use $N_f=1024$ subcarriers and truncate the delay-domain CSI matrix to include the first $R_d=32$ rows.

\begin{figure*}[!hbtp] \centering 
	\subfigure[Indoor] {\label{fig:slot1_indoor} 
	\includegraphics[width=0.46\textwidth]{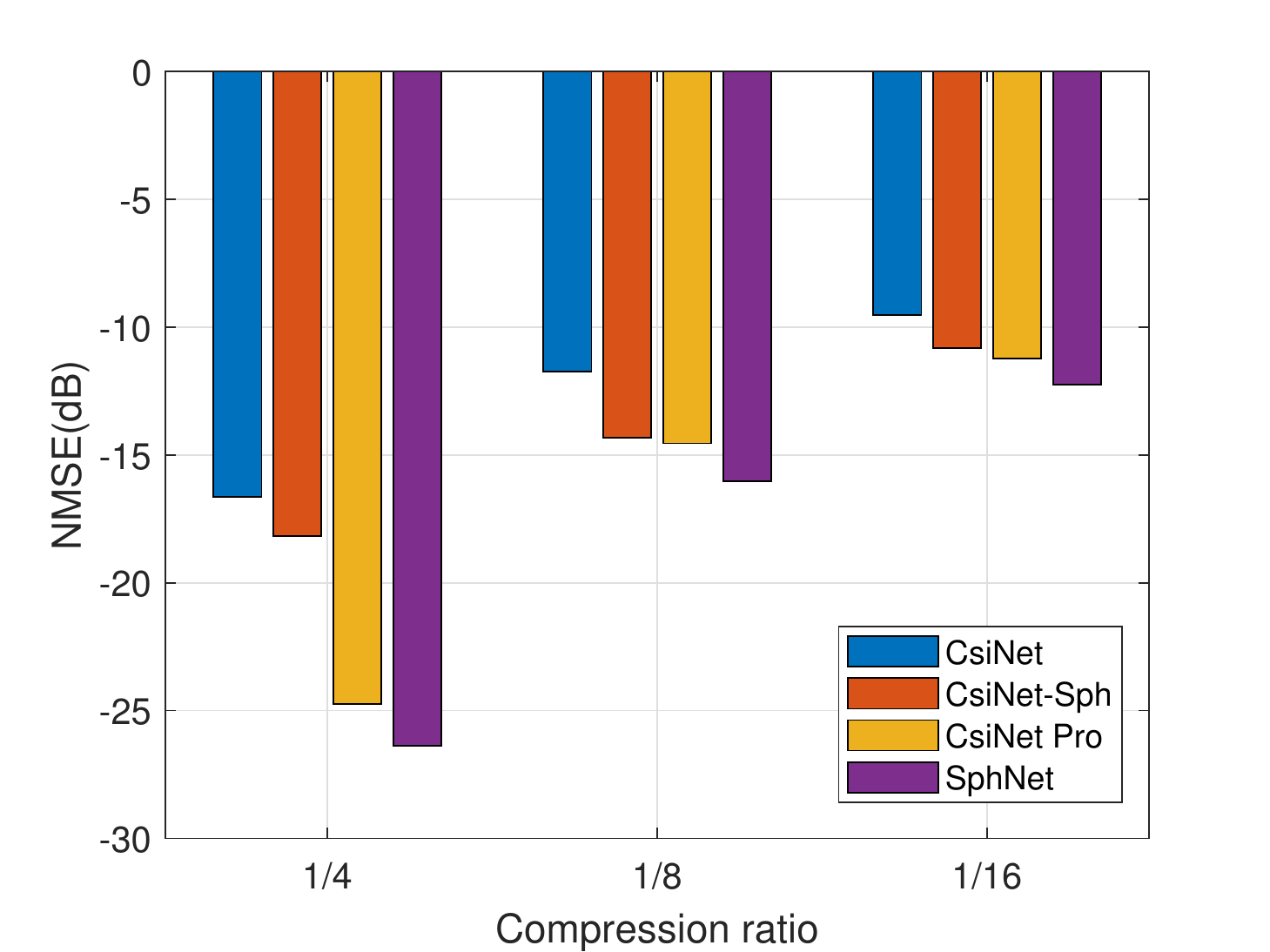}
	} 
	\subfigure[Outdoor] { \label{fig:slot1_outdoor} 
	\includegraphics[width=0.46\textwidth]{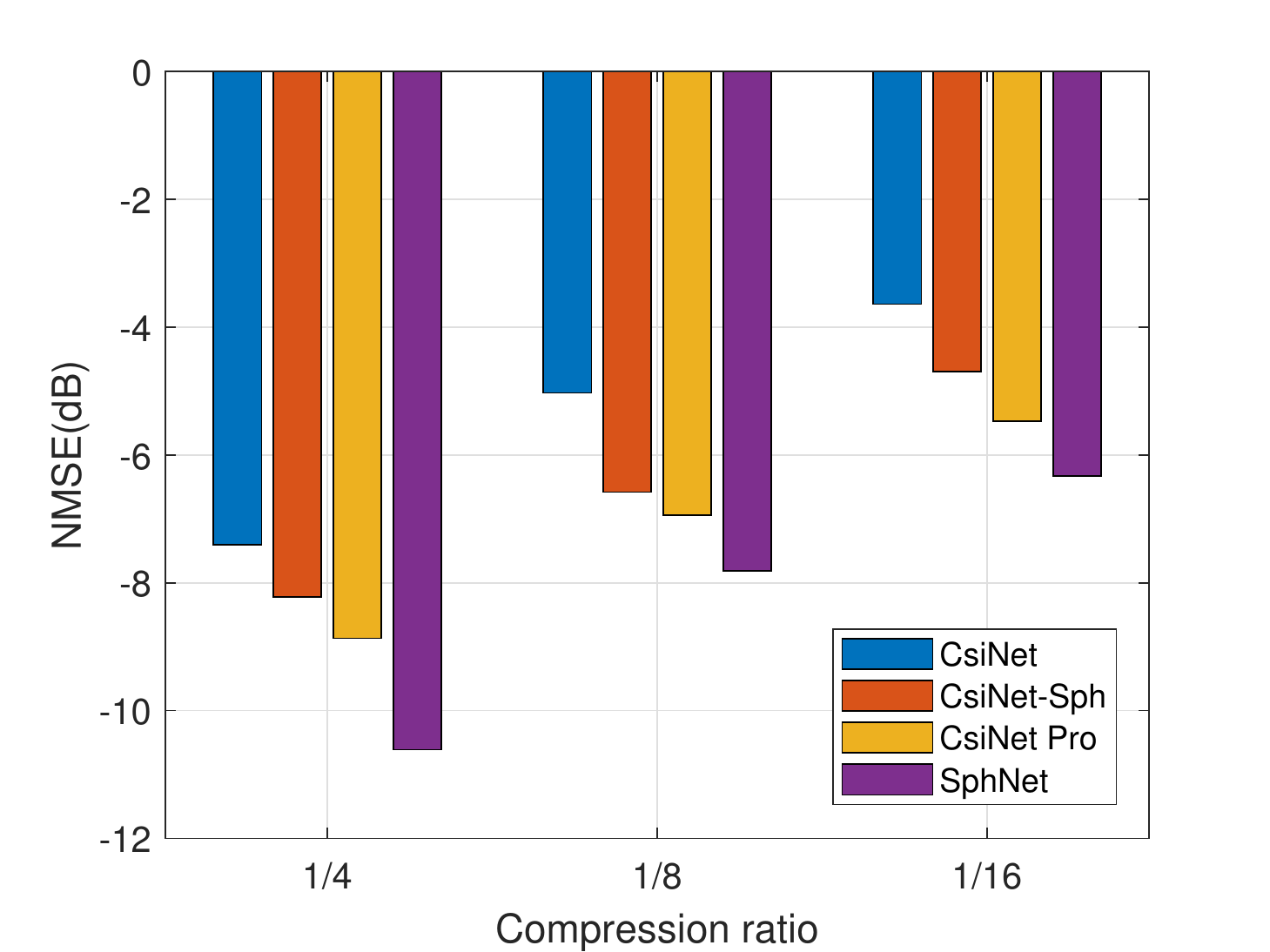} 
	} 
	\caption{NMSE of different networks in the first time slot of MarkovNet over varying compression ratios (CR).} 
	\label{fig:nmse_slot1} \vspace*{-2mm}
\end{figure*}

\begin{figure*}[!hbtp] \centering 
	\subfigure[MarkovNet and CsiNet-LSTM indoor] {\label{fig:diffnet_indoor} 
	\includegraphics[width=0.46\textwidth]{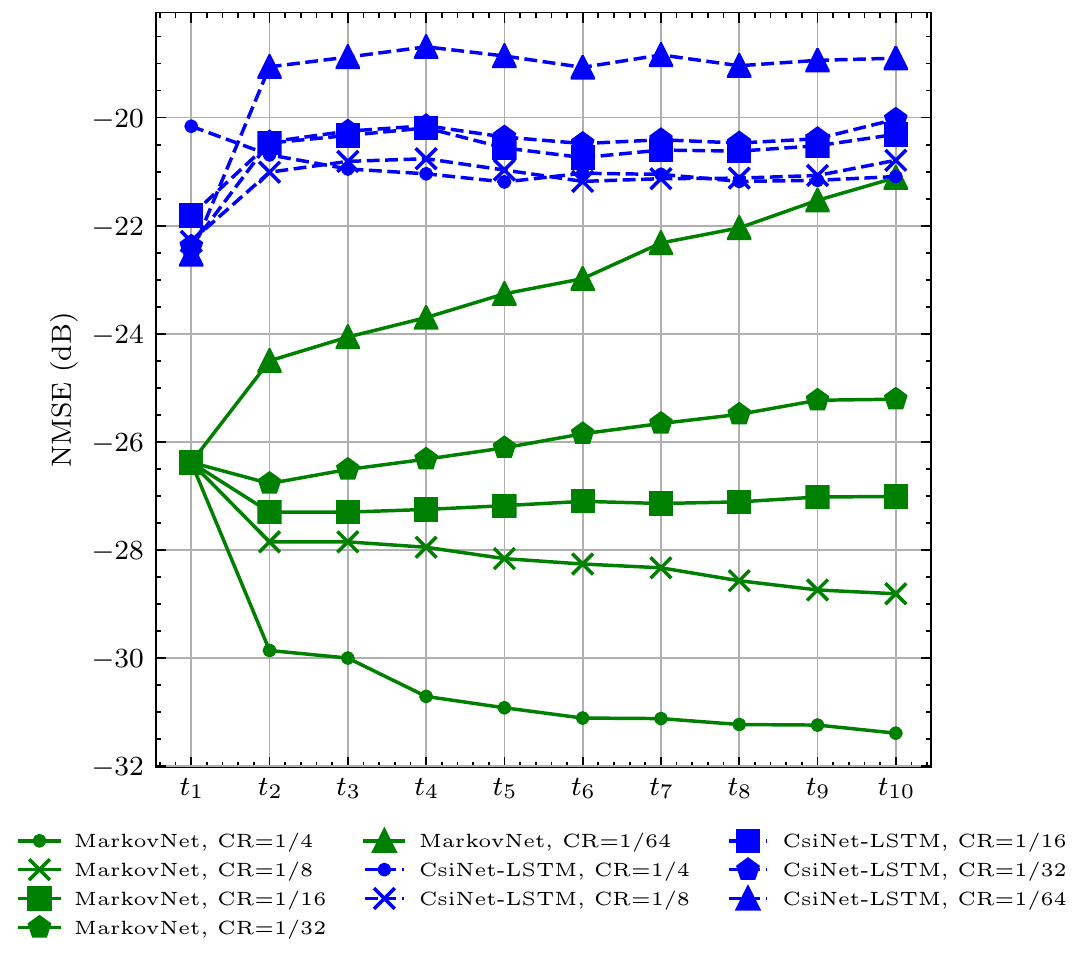}
	} 
	\subfigure[MarkovNet and CsiNet-LSTM outdoor] { \label{fig:diffnet_outdoor} 
	\includegraphics[width=0.46\textwidth]{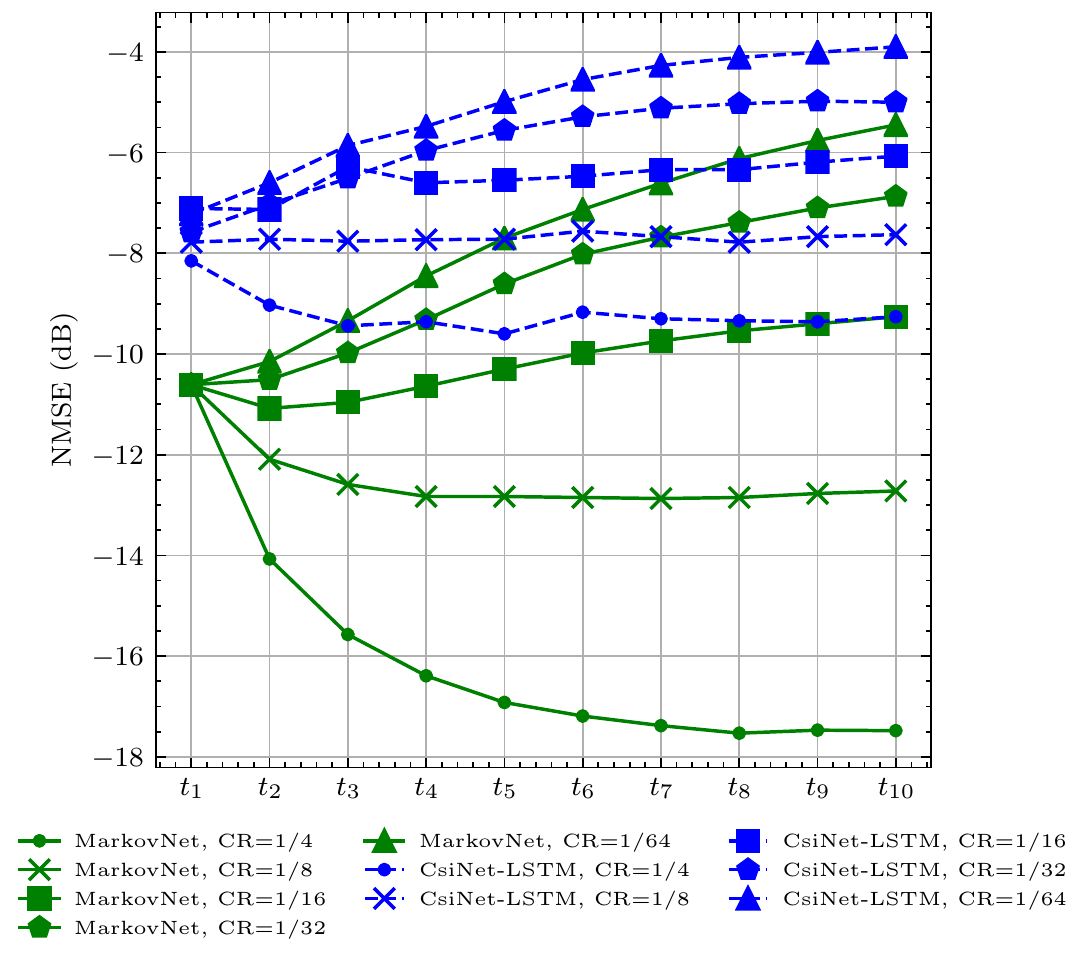} 
	} 
	\caption{NMSE comparison between MarkovNet and CsiNet-LSTM over varying compression ratios (CR).} 
	\label{fig:diffnet_result} \vspace*{-2mm}
\end{figure*}

The gNB uses antennas arranged in a uniform linear array (ULA) with half-wavelength spacing.  UEs are randomly positioned within
the coverage area such that their CSIs are random. 
For each indoor/outdoor environment, we generate a dataset of $10^5$ 
sample channels and divide them
into $7.5\cdot 10^4$ and $2.5\cdot 10^4$
for training and testing sets, respectively. The batch size for the training of MarkovNet is 200.
MarkovNet at t1 was trained for 1000 epochs using MSE 
as the loss function. 
For the MarkovNet after $t_2$, 
only 150 epochs are used with the help of initialization 
using the pretrained model of the previous time slot 
to reduce training expenses. We utilize the Adam optimizer with 
default learning rate $10^{-3},$  and hyperparameters (i.e., batch size, epochs) for each test will be clarified in each relevant subsection. 
NMSE is used to compare the CSI recovery accuracy of different networks.
% For training, we utilize Adam with a learning rate of $10^{-3}$ and a batch size of 200. 
% Each network was trained for 1000 epochs using MSE (as Eq. (\ref{eq:mse})) as the loss function, and the figure of merit used to compare all networks was NMSE (as Eq. (\ref{NMSE})).

% \subsection{CSI Feedback at Initial Time Slot}

% We note that further improvement of feedback efficiency is possible by using well defined feedback quantizers as encoders. 

\subsection{MarkovNet}
In this part, 
we evaluate the performance of MarkovNet 
considering the performance at the first timeslot ($t_1$), 
the overall performance of MarkovNet, and the performance of MarkovNet-CNN.

\subsubsection{Performance evaluation at $t_1$}
% We demonstrate substantial error performance gains over methods using standard. 
To enable efficient differential CSI feedback, high accuracy CSI feedback is required at $t_1$ to provide
a good starting CSI condition for subsequent timeslots. 
Here, we demonstrate that our proposed spherical CSI feedback framework improves
the CSI recovery accuracy for a single time slot 
compared to different CSI feedback frameworks.

\begin{figure*}[!hbtp] \centering 
	\subfigure[indoor] {\label{fig:diffnet_cnn_indoor} 
	\includegraphics[width=0.46\textwidth]{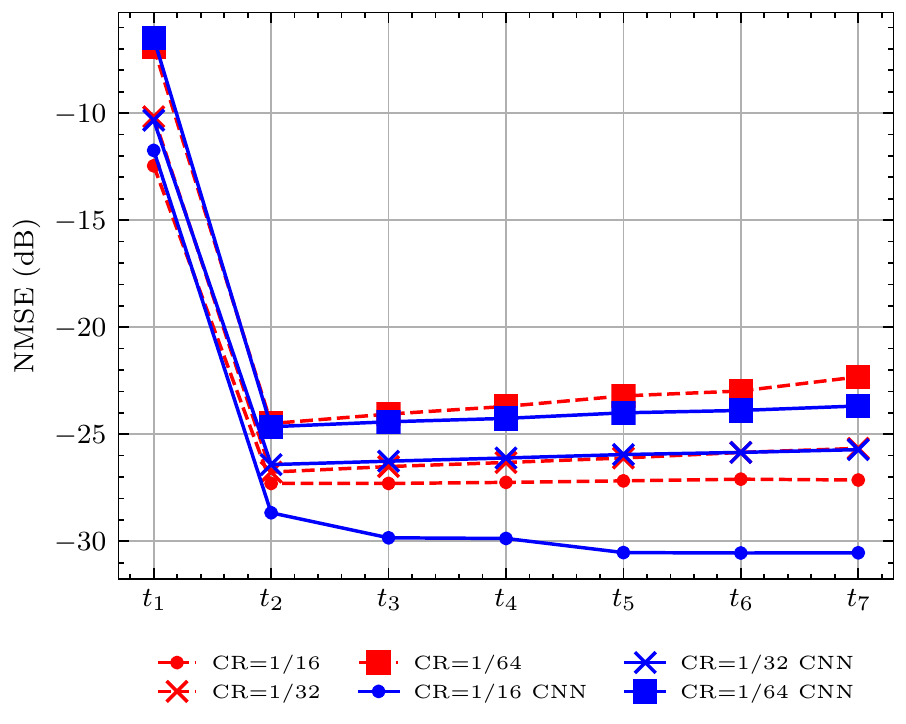}
	} 
	\subfigure[outdoor] { \label{fig:diffnet_cnn_outdoor} 
	\includegraphics[width=0.46\textwidth]{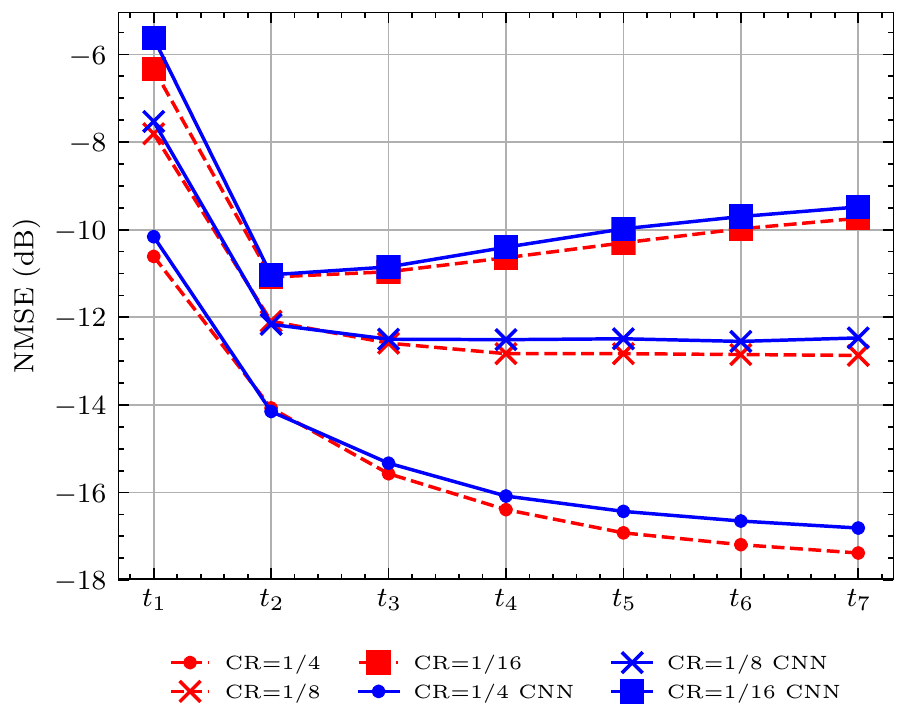} 
	} 
	\caption{NMSE comparison between MarkovNet and MarkovNet-CNN over varying CR.} 
	\label{fig:diffnet_cnn} \vspace*{-2mm}
\end{figure*}

Fig. \ref{fig:nmse_slot1} compares
the performance of channel reconstruction
from the use of CsiNet, CsiNet-Sph 
(CsiNet with the help of spherical feedback), 
CsiNet Pro, and 
SphNet (CsiNet Pro with the help of the spherical feedback framework). 
As shown in Fig. \ref{fig:nmse_slot1}, 
SphNet achieves the best performance
in single shot feedback for CSI recovery without relying on
prior CSI knowledge, which means that SphNet can improve the accuracy of prior information for the MarkovNet. On the one hand, CsiNet Pro outperforms the CsiNet in different CR and scenarios, which means the enhanced network structure is effective.
% Having a strong starting CSI recovery, this also would imply 
% that SphNet can provide the best CSI
% reconstruction in subsequent time slots
% of MarkovNet. 
On the other hand, we can observe that spherical feedback can provide the 
most noticeable performance gain to both CsiNet and CsiNet Pro.
This establishes the strength of spherical normalization
to efficiently capture the CSI data feature. 

\subsubsection{Overall performance evaluation of MarkovNet}

Every instance of MarkovNet contains two different compression ratios.
For the first time slot, we initialize MarkovNet with CR=$1/4$
at $t_1$ to provide an accurate CSI baseline for subsequent time slots.
For the rest timeslots, MarkovNet maintains 
the same CR for all subsequent timeslots ($t_2$ to $t_{10}$).
To evaluate MarkovNet's performance under different amounts of compression,
we vary the second CR used in the later timeslots 
from $1/4$ to $1/64$ and train each network. 
For example, in the Fig. \ref{fig:diffnet_result} that follows, ``MarkovNet, CR=1/16" 
uses CR=1/16 at timeslots $t_2$ through $t_{10}$ 
and CR=1/4 at timeslot $t_1$.

Fig. \ref{fig:diffnet_result} compares the performance between 
MarkovNet and CsiNet-LSTM. 
The benefit of differential CSI encoding is demonstrated by
the CSI recovery accuracy of MarkovNet using different compression
ratios beyond $t_2$ in comparison with CsiNet-LSTM. 
MarkovNet consistently achieves higher
CSI accuracy than CsiNet-LSTM at every CR level.
With the help of differential CSI encoding,
MarkovNet is an effective encoding framework given
limited UE power and bandwidth for CSI encoding. 
For the indoor channels, MarkovNet can deliver 
reliable CSI accuracy of -30dB even for the compression ratio of
$1/64$, which is a 10dB improvement over CsiNet-LSTM. 
Although the outdoor scenario continues to be more challenging,
our results show that 1/4 or 1/8 compression ratio can
achieve NMSE of $-17$dB and $-12$dB, respectively. On the
other hand, CsiNet-LSTM is shown to
provide NMSE only at $-9$ and $-7.5$dB, respectively. 

\subsubsection{Performance and Complexity Trade-off of MarkovNet-CNN}
Fig. \ref{fig:diffnet_cnn} shows the performance comparison 
between the MarkovNet and MarkovNet-CNN at
different meaningful compression ratios. Since the trend of CSI
accuracy is similar over time, we focus on the performance 
from $t_1$ to $t_7$. 
For the first time slot, we initialize MarkovNet and MarkovNet-CNN with CR=$1/4$
at $t_1$ to provide an accurate CSI baseline for subsequent time slots. Note that, to show the influence of CNN-based dimension compression module at $t_1$, the results we shown in Fig. \ref{fig:diffnet_cnn} at $t_1$ are from the labeled compression ratios.
Both
MarkovNet and MarkovNet-CNN achieve comparable CSI accuracy at $t_1$,
indicating that CNN layer for compression and decompression is
not only more efficient in memory and computation, but
also delivers similar CSI accuracy.  For the rest timeslots, MarkovNet maintains 
the same CR for all subsequent timeslots ($t_2$ to $t_{7}$).
Beyond $t_2$, MarkovNet-CNN achieves modestly higher accuracy 
for indoor channels as shown in Fig. \ref{fig:diffnet_cnn}(a). 
The benefit of MarkovNet-CNN likely arises from the
reduction of many redundant weights from the FC layer
such that there are fewer opportunities for local minimum
convergence. For outdoor channels,
MarkovNet-CNN achieves comparable CSI accuracy as MarkovNet 
for compression ratio of $1/8$ and $1/16$ 
while exhibiting a modest loss of accuracy at CR=1/4.
One possible reason is that outdoor channels can
benefit more from higher number of connectivity in layers for
compression and feature extraction because of their much
more complex characteristics. 

\subsection{Model size and Computational Complexity}

We demonstrate that latent convolutional layers 
require
significantly fewer parameters than FC-layers 
without loss of performance. Table~\ref{tab:comp-complex}
compare the model size and computational 
complexity (respectively) of CsiNet-LSTM, MarkovNet, and MarkovNet-CNN associated with a single timeslot. 
Across all compression ratios,
MarkovNet uses 60 times fewer parameters than CsiNet-LSTM. 
More importantly, MarkovNet-CNN can use 1/3000
the number of parameters needed by CsiNet-LSTM 
while achieving better CSI recovery accuracy.

\begin{figure*}[!hbtp] \centering 
	\subfigure[Sweep depth for quantized/perfect CSI (Indoor)] {\label{fig:lstm-sweep-in} 
	\includegraphics[width=0.46\textwidth]{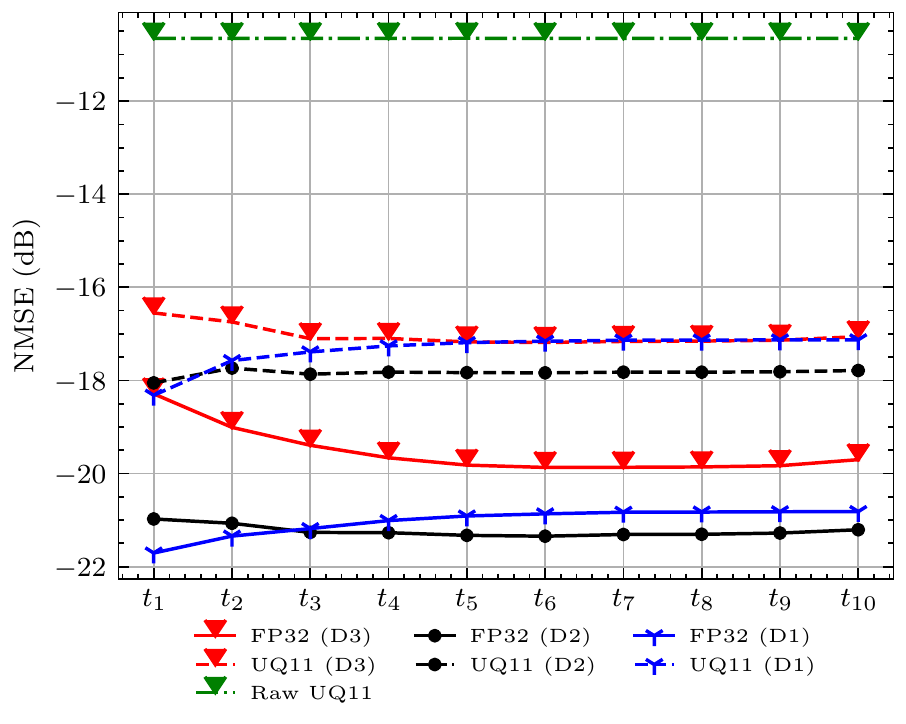} 
	} 
	\subfigure[Sweep depth for quantized/perfect CSI (Outdoor)] {\label{fig:lstm-sweep-out} 
	\includegraphics[width=0.46\textwidth]{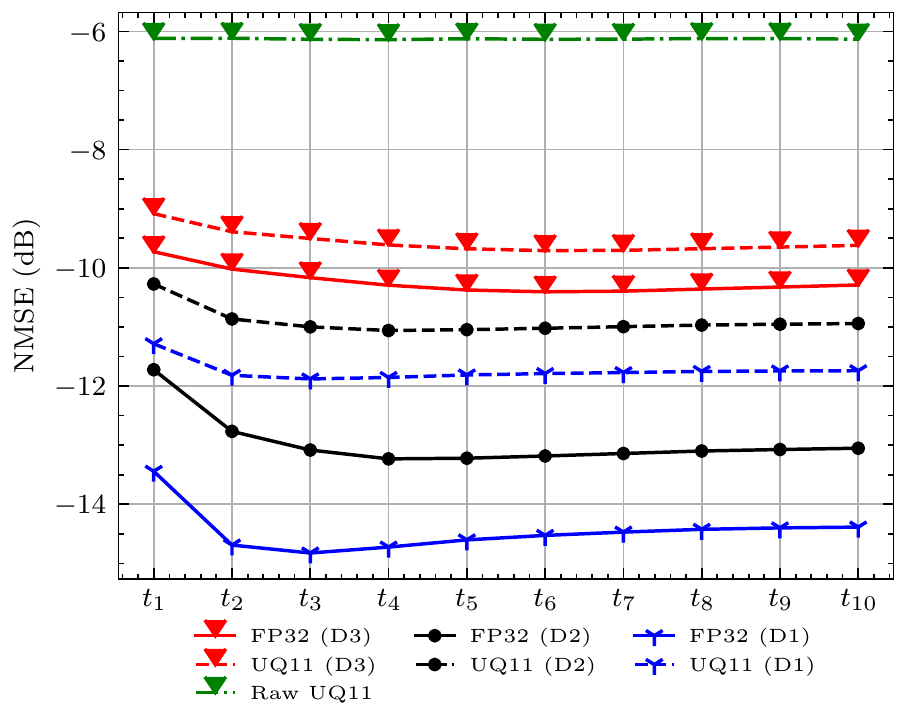} 
	} 
	\caption{Stacked LSTMs for 3 depths trained on perfect and quantized CSI trained on Indoor 5.3GHz network (a) and on Outdoor 300MHz network (b).} 
	\label{fig:lstm-vary} \vspace*{-2mm}.
\end{figure*}

\begin{table*}[]
\renewcommand{\arraystretch}{1.5}
\centering
\caption{Model size and computational complexity of tested networks. M: million, K: thousand.}
\label{tab:comp-complex}
\begin{tabular}{|c|c|c|c|c|c|c|}
\hline
% & \multicolumn{1}{c|}{\textbf{CsiNet-LSTM}} & \multicolumn{1}{c|}{\textbf{MarkovNet}} & \multicolumn{1}{c|}{\textbf{MarkovNet-CNN}} \\ \hline
 & \multicolumn{3}{c|}{\textbf{Parameters}} & \multicolumn{3}{c|}{\textbf{FLOPs}} \\ \hline
 & \textbf{CsiNet-LSTM} & \textbf{MarkovNet} & \textbf{MarkovNet-CNN} & \textbf{CsiNet-LSTM} & \textbf{MarkovNet} & \textbf{MarkovNet-CNN} \\ \hline
\textbf{CR=$\frac {1}{4}$} & 132.7 M         & 2.1 M        & 34.9 K     &  412.9 M            & 44.5 M         & 41.2 M         \\ \hline
\textbf{CR=$\frac {1}{8}$} & 123.2 M         & 1.1 M      & 27.8 K       &  410.8 M            & 42.4 M          & 40.7 M         \\ \hline
\textbf{CR=$\frac {1}{16}$}& 118.5 M      & 542.9 K         & 24.2 K     &  409.8 M            & 41.3 M         & 40.5 M         \\ \hline
\textbf{CR=$\frac {1}{32}$}& 116.1 M         & 280.7 K         & 22.4 K  &  409.2 M           & 40.8 M         & 40.4 M         \\ \hline
\textbf{CR=$\frac {1}{64}$}& 115.0 M         & 149.6 K          & 21.5 K &  409.0 M            & 40.5 M         & 40.3 M       \\ \hline
\end{tabular}
\end{table*}

Table~\ref{tab:comp-complex} also shows the average
number of floating point operations (FLOPs) associated with a single timeslot for each network \cite{ref:Molchanov2016,ref:Nisar2019}. MarkovNet and MarkovNet-CNN can 
save computation load by more than $\frac{8}{9}$ and $\frac{9}{10}$ FLOPs, respectively, when compared with the CsiNet-LSTM in each compression ratio.  
For CsiNet-LSTM, the amount of computation does not 
change significantly even at low compression ratios.
For example, a 16-fold drop in compression ratio
(from 1/4 to 1/64) only 
results in a 1\% saving of FLOPs. 
In contrast, MarkovNet and MarkovNet-CNN require much lower
computational complexity in proportion at lower compression
ratios.  In MarkovNet and MarkovNet-CNN networks, for example,
a 16-fold CR reduction (from 1/4 to 1/64)
reduces the number of FLOPs by 9\% and 2\%,
respectively.

We note that when deploying MarkovNet and MarkovNet-CNN
as a cooperative learning mechanism at both UE and
gNB, 50\% additional parameters and FLOPs are required 
in comparison with the training phase. This is because 
the trained decoder must be duplicated at the UE side
to generate the decoded CSI for the previous time slot
used by the encoder. Despite this additional cost,
both MarkovNet and MarkovNet-CNN still can reduce the
number of parameters by orders of magnitude, 
and save over $\frac{5}{6}$ FLOPs in comparison 
with CsiNet-LSTM.

\subsection{LSTMs for CSI Estimation} \label{sec:rnn-for-csi}

\begin{table}[!hbtp]
\renewcommand{\arraystretch}{1.5}
\caption{Average NMSE across ten timeslots ($T=10$) for stacked LSTMs of increasing depth trained on quantized CSI under 11-bit uniform quantization (`UQ11' in Fig.~\ref{fig:lstm-vary}). NMSE of quantized CSI under uniform quantization (`Raw UQ11') is shown for comparison.}
\begin{center}
\begin{tabular}{|c|c|c|c|c|}
\hline
\textbf{Environment} & \textbf{Raw UQ11} & \textbf{Depth 1} & \textbf{Depth 2} & \textbf{Depth 3} \\ \hline
Indoor      &  -10.66 dB   & -21.03 dB  & -21.26 dB  & -19.54 dB  \\ \hline
Outdoor     &   -6.41 dB  & -14.64 dB & -13.11 dB   & -10.40 dB   \\ \hline
\end{tabular}
\end{center}
\label{tab:lstm-quant}
\end{table}

In this section, we explore the effects of varying LSTM depth on network performance. For all experiments, we use the Adam optimizer with learning rate of $10^{-3}$ and a batch size of 100. For the LSTM-only networks in the ablation study, we train for 500 epochs. For CsiNet-LSTM, we pretrain CsiNet at each compression ratio for 600 epochs, and we then initialize the CsiNet at each timeslot of CsiNet-LSTM with the pretrained weights before training for 500 epochs.

\subsubsection{Ablation study on LSTM depth}

% place s.t. this is at top of col
% \begin{table}[!hbtp]
% \renewcommand{\arraystretch}{1.5}
% \caption{Average NMSE across ten timeslots ($T=10$) for CsiNet-LSTM \cite{ref:csinet-lstm} with two different LSTM depths, 1 and 3. Trained for 500 epochs.}
% \begin{center}
% \begin{tabular}{|c|c|c|c|c|}
% \hline
% \textbf{Environment}     & \textbf{Depth} & \textbf{CR=$\frac {1}{16}$} & \textbf{CR=$\frac {1}{32}$} & \textbf{CR=$\frac {1}{64}$} \\ \hline
% \multirow{2}{*}{Indoor}  & 3              &  -22.86                     & -22.94                      & -19.97                    \\ \cline{2-5} 
%                          & 1              &  -24.01                     & -23.28                      & -19.48 \\ \hline
% \multirow{2}{*}{Outdoor} & 3              &  -6.62                      & -5.85                       & -5.09 \\ \cline{2-5} 
%                          & 1              &  -6.99                      & -6.08                       & -5.35\\ \hline
% \end{tabular}
% \end{center}
% \label{tab:csinet-lstm}
% \end{table}

We seek to know whether shallow RNNs perform comparably well to deep networks. To investigate the effect of network depth, we train stacked LSTMs of increasing depth on $\{\bar{\mathbf{H}}_t\}_{t=1}^{10}$, which are CSI matrices quantized with two different schemes: 1) Single-precision floating point (FP32) and 2) 11-bit Uniform Quantization (UQ11). For a visual depiction of this network, see Fig.~\ref{fig:lstm-example}(b). We train each network with Adam using the default

% we trained LSTM layers on `quantized' CSI, $\{\mathbf{H}^{q}_t\}_{t=1}^T$, and `perfect' CSI, $\{\mathbf{H}_t\}_{t=1}^T$. The perfect CSI, $\mathbf{H}_t$, is normalized to the range $[-1,1]$, and the quantized CSI, $\mathbf{H}^{q}_t$, at timeslot $t$ is the perfect CSI quantized to $2^b$ levels with $b$ bits. This quantization introduces noise to the CSI samples with no compression. % rather than CSI estimates $\{\hat{\mathbf{H}}_t\}_{t=1}^T$.

\begin{figure*}[!hbtp] \centering 
%     \subfigure[Indoor]{
%     	\includegraphics[width=0.46\linewidth]{Indoor_all.pdf}\label{fig:csinetlstm-in-d-comp}
% 	}
% 	\subfigure[Outdoor]{
%     	\includegraphics[width=0.46\linewidth]{Outdoor_all.pdf}\label{fig:csinetlstm-out-d-comp}
% 	}
    \includegraphics[width=\linewidth]{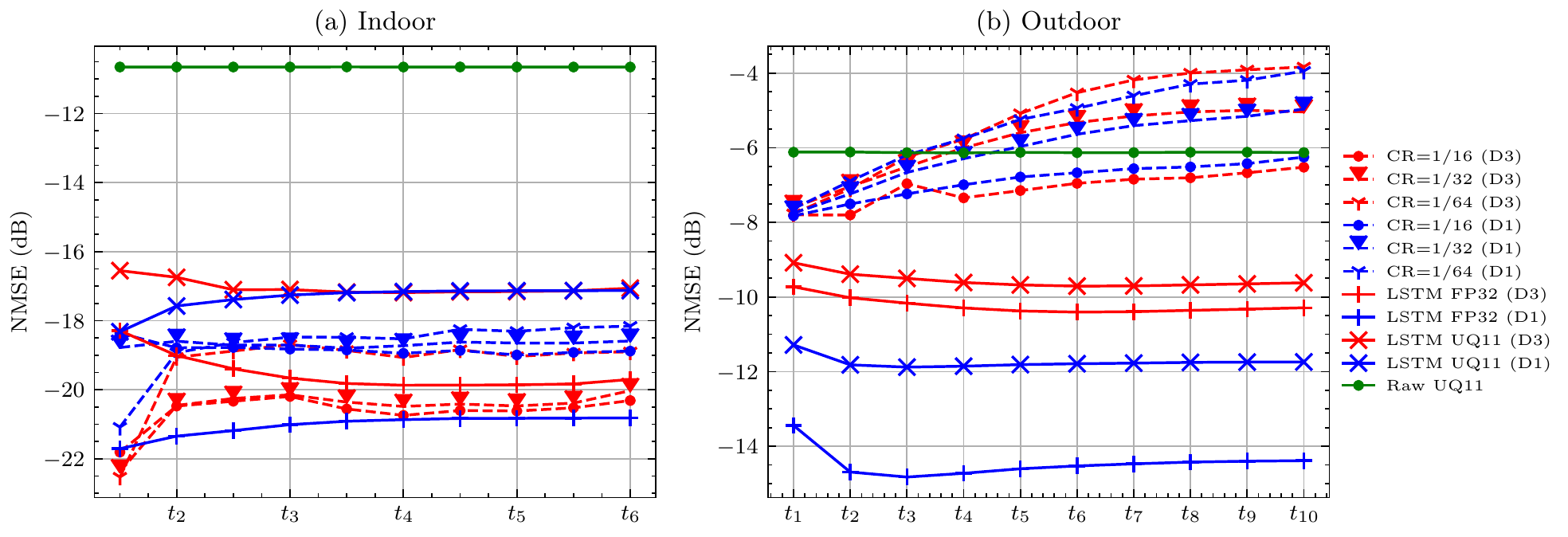}
	\caption{CsiNet-LSTM over varying compression ratios (CR) compared to LSTMs trained on perfect CSI and quantized CSI. D$N$ indicates a stacked LSTM depth $N$.}
	\label{fig:csinetlstm-d-comp} \vspace*{-2mm}
\end{figure*}

 Figure~\ref{fig:lstm-vary} shows the NMSE per timeslot for each of these RNNs, and the average performance across all timeslots is shown in Table~\ref{tab:lstm-quant}. At all depths, LSTMs are able to improve the test NMSE relative to quantized CSI. However, there is not a clear linear relationship between network depth and NMSE performance. For the outdoor network (Fig.~\ref{fig:lstm-sweep-out}), the network performance and LSTM depth appear negatively correlated -- increasing depth results in decreasing performance. For the indoor network, the best network has a depth of $D=2$, indicating the the best choice of depth is channel-dependent.

% \begin{table}[htb]
% \centering
% \renewcommand{\arraystretch}{1.5}
% \caption{Conditional entropy, $H(\mathbf{H}_{t}|\mathbf{H}_{t-t_0})$, (bits/element) under different speeds and feedback intervals ($t_0$). The baseline corresponds to the entropy without prior information, $H(\mathbf{H}_{t})$.}
% \begin{tabular}{|c|c|c|c|c|c|}
% \hline
%                         & \textbf{baseline} & \textbf{40ms} & \textbf{80ms} & \textbf{160ms} & \textbf{320ms} \\ \hline
% \textbf{outdoor 0.9m/s} & 4.77              & 2.29          & 2.81          & 3.64           & 4.41           \\ \hline
%                         & \textbf{baseline} & \textbf{5ms}  & \textbf{10ms} & \textbf{20ms}  & \textbf{40ms}  \\ \hline
% \textbf{outdoor 3.6m/s} & 4.61              & 1.53          & 2.03          & 2.65           & 3.42           \\ \hline
% \end{tabular}
% \label{tab:conditional_entropy}
% \end{table}

\subsubsection{LSTM Depth in CsiNet-LSTM}

While LSTMs can perform admirably when using noisy CSI samples, these samples were not subject to compression. Compression is imperative for channel feedback, as transmitting uncompressed CSI will consume an undue amount bandwidth.

In this experiment, we use a known CNN/RNN for CSI estimation, CsiNet-LSTM \cite{ref:csinet-lstm}. Fig.~\ref{fig:csinetlstm-d-comp} illustrates
the NMSE for different depths of CsiNet-LSTM in each of the 10 time slots.  The original network utilizes LSTMs of depth 3 (D3 in Figures ~\ref{fig:csinetlstm-d-comp}(a), \ref{fig:csinetlstm-d-comp}(b)), and we also
 train CsiNet-LSTM with one LSTM layer (D1) for comparison. 
We show the performance of the D3 and D1 networks to 
LSTMs trained directly 
on FP32 and UQ11 CSI samples (i.e., the same networks in Fig.~\ref{fig:lstm-vary}. We train each network end-to-end based on the original paper's hyperparameters and dataset splits.

Figure~\ref{fig:csinetlstm-d-comp}(a) shows  
the NMSE in channel reconstruction
by CsiNet-LSTM for the indoor dataset.
While the shallower D1 network with fewer parameters
in fact outperformed the deeper D3 network 
in the FP32 and UQ11 scenarios, 
the D3 variant of CsiNet-LSTM performs better than the D1 version
for the all three compression ratios.

This performance trend relative to LSTM depth does not hold
for the outdoor network.
Figure~\ref{fig:csinetlstm-d-comp}(b) shows 
 the NMSE in channel reconstruction for the outdoor dataset. 
Clearly, the shallower
D1 network performs similarly to the D3 network at
the tested compression ratios.

These results indicate that a simple, shallower LSTM 
can perform similarly to complex, deeper
networks for both indoor and outdoor datasets.

\begin{table}[!hbtp]
\renewcommand{\arraystretch}{1.5}
\caption{MarkovNet and CsiNet-LSTM mean NMSE degradation (increase) under different feedback quantization bits. The mean is taken across all tested compression ratios, and the degradation in NMSE is relative to floating point 32 bit precision.}
\begin{center}
\begin{tabular}{|c|c|c|c|}
\hline
\textbf{Network}                           & \textbf{Environment} & \textbf{6 bits} & \textbf{4 bits} \\ \hline
\multirow{2}{*}{\textbf{MarkovNet}}          & Indoor      &  0.70 dB  &  5.49 dB \\ \cline{2-4}
                                           & Outdoor     &  0.03 dB  &  0.58 dB \\ \hline
\multirow{2}{*}{\textbf{CsiNet-LSTM (D3)}}  & Indoor     &  2.30 dB  & 11.96 dB \\ \cline{2-4}
                                           & Outdoor     &  0.07 dB  &  1.25 dB  \\ \hline
\multirow{2}{*}{\textbf{CsiNet-LSTM (D1)}} & Indoor      &  2.44 dB  & 11.55 dB \\ \cline{2-4}
                                           & Outdoor     &  0.30 dB  &  1.21 dB  \\ \hline
\end{tabular}
\end{center}
\label{tab:quant-mean}
\end{table}

\subsection{Network Performance Under Feedback Quantization}

To understand the effect of feedback quantization, we apply $\mu$-law companding to the encoded layer of both tested networks. $\mu$-Law companding uses a logarithmic transformation that emphasizes lower magnitude samples. For signal value $x$, the compression portion of the $\mu$-law scheme is written as 
\begin{align}
    f(x) &= \frac{\text{sgn}(x)\ln (1+\mu|x|)}{\ln (1+\mu)}, \; 0 \leq |x| \leq 1. \label{eq:mu-compress}
\end{align}
Uniform quantization is applied to the compressed signal. For signal value $x$, the quantization/dequantization operation produces a value $\hat x$, which can be written
\begin{align}
    \hat f(x) &= \Delta \left\lfloor\frac{f(x)}{\Delta}\right\rceil \label{eq:uni-quant}
\end{align}
for fixed step size $\Delta$. After the quantized feedback is received, then we expand the result using the inverse of (\ref{eq:mu-compress}),
\begin{align}
    % F(\hat x) &= \text{sgn}(\hat x)(1 / \mu) ((1+\mu)^{|\hat x|} - 1), \; -1 \leq y \leq 1. \label{eq:mu-expand}
    F(\hat x) &= \frac{\text{sgn}(\hat f(x)) ((1+\mu)^{|\hat f(x)|} - 1)}{\mu}, \; -1 \leq y \leq 1. \label{eq:mu-expand}
\end{align}

\begin{figure*}[!hbtp] \centering 
	\includegraphics[width=0.95\textwidth]{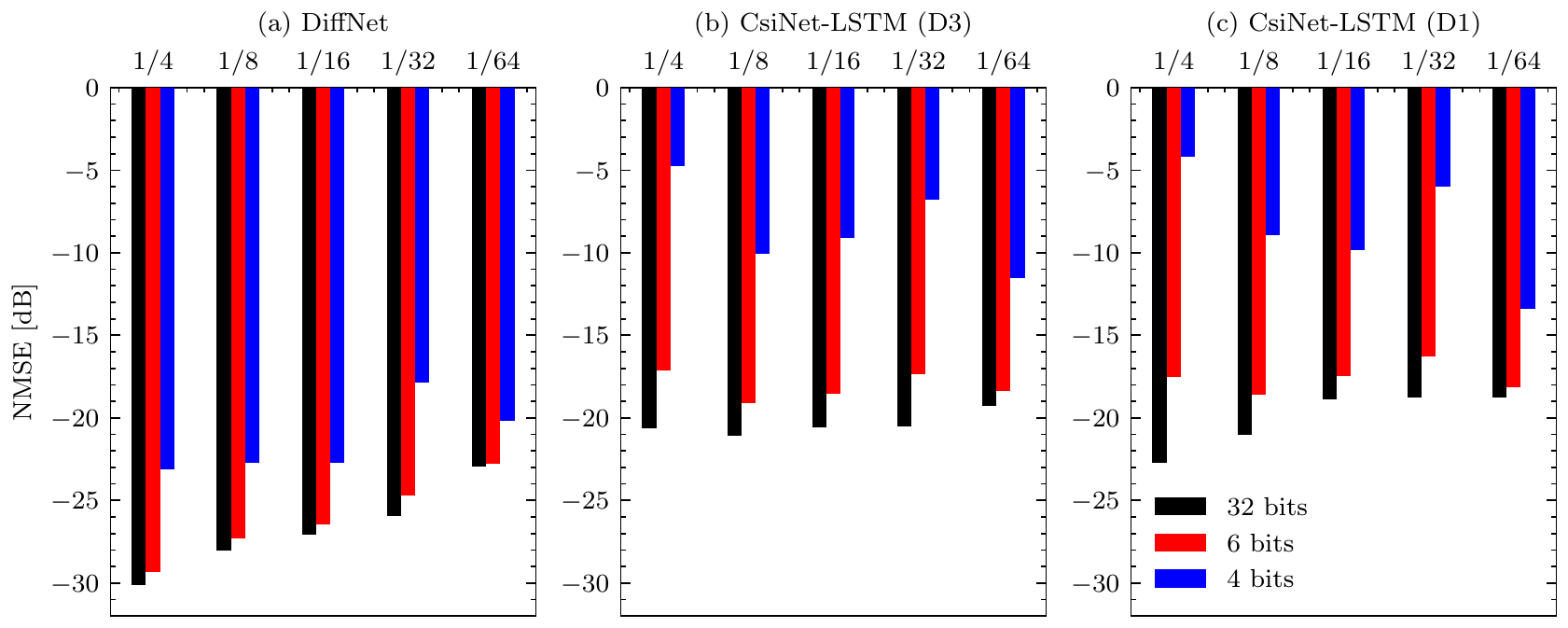}
    \caption{MarkovNet and CsiNet-LSTM NMSE performance (dB) for indoor network with feedback subject to mu-law quantization using fixed step size, $\Delta=2^{b-1}$, for $b$ bits.}
	\label{fig:feedback_quant_indoor} \vspace*{-2mm}
\end{figure*}

\begin{figure*}[!hbtp] \centering 
	\includegraphics[width=0.95\textwidth]{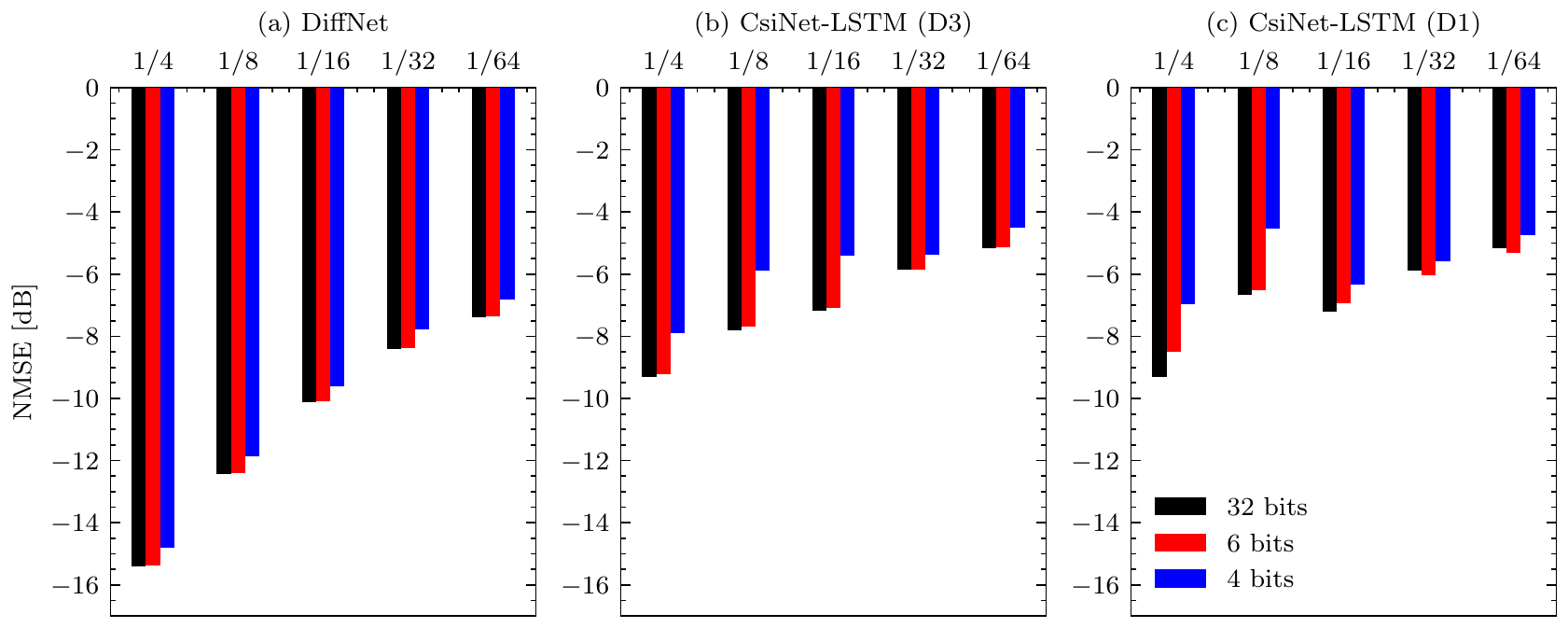}
    \caption{MarkovNet and CsiNet-LSTM NMSE performance (dB) for Outdoor network with feedback subject to mu-law quantization using fixed step size, $\Delta=2^{b-1}$, for $b$ bits.}
	\label{fig:feedback_quant_outdoor} \vspace*{-2mm}
\end{figure*}

Fig.~\ref{fig:feedback_quant_indoor} and Fig.~\ref{fig:feedback_quant_outdoor} show the performance of MarkovNet and CsiNet-LSTM with $\mu$-law companding and fixed quantization step size at two different quantization levels, 6 bits and 4 bits, in comparison to the non-quantized network (i.e., 32 bit floating point). The networks with quantized feedback use 8 bit quantization at the first timeslot to establish good intial CSI estimates. Note that the networks are not re-trained or fine-tuned after applying quantization.

MarkovNet is more robust to feedback quantization noise than CsiNet-LSTM at either LSTM depth, as the former maintains NMSE better than -10 dB in both environments while the latter only exceeds -10dB at low compression ratios in the Indoor environment. Table~\ref{tab:quant-mean} summarizes the mean decrease in NMSE for each network, and for 6 and 4 bit $\mu$-law quantization, MarkovNet has a smaller mean degradation in NMSE performance compared to CsiNet-LSTM.

\section{Conclusion} \label{sec:conclusion}

 To better exploit temporal correlation, we provide an information theoretic basis for utilizing differential encoding with CNNs rather than applying overly parameterized LSTMs. We propose MarkovNet, a CNN with differential encoding, which achieves superior estimation accuracy and lowers computational complexity relative to an LSTM-based CSI estimation network. MarkovNet maintains accurate CSI estimates even under feedback quantization. We expand on a prior LSTM-based estimation technique and show that more parsimonious models can yield comparable or better performance. 

\section*{Acknowledgment}
% The authors wish to thank Prof. S. Jin and his coauthors 
% of \cite{ref:csinet-lstm} for providing source codes of benchmark algorithms.

The authors wish to thank Prof. S. Jin of Southeast University for his kind assistance on source codes of \cite{ref:csinet-lstm} and related questions
in the process of preparing for this manuscript.

\bibliographystyle{IEEEtran}
\bibliography{markov_net.bib}

\end{document}